\documentclass[journal=jpc,manuscript=article,layout=twocolumn,superscriptaddress,amsmath,amssymb,aps,floatfix]{achemso}
\usepackage[version=3]{mhchem} % Formula subscripts using \ce{}
\usepackage{url}
\usepackage{ulem}
\usepackage{makeidx}
\usepackage{color}
\usepackage{graphicx}
\usepackage{subcaption}
\usepackage{verbatim}
\usepackage{bibunits}
\makeindex
\SectionNumbersOn
\newcommand{\e}[1]{{\rm e}^{#1}}

\author{Peter E. Bl\"ochl}
\affiliation{Institute for Theoretical Physics, Clausthal University of Technology, Germany}
\alsoaffiliation{Center for Advanced Systems Understanding (CASUS), Helmholtz Zentrum Dresden-Rossendorf, Germany}
\alsoaffiliation{Institute for Theoretical Physics, Georg-August University, G\"ottingen, Germany}
\email{peter.bloechl@tu-clausthal.de}
\author{Robert Schade}
\affiliation{Paderborn Center for Parallel Computing (PC2), Paderborn University, Germany}
\author{Lukas Allen-Rump}
\affiliation{Institute for Theoretical Physics, Clausthal University of Technology, Germany}
\alsoaffiliation{Institute for Theoretical Physics, Georg-August University, G\"ottingen, Germany}
\author{Sangeeta Rajpurohit}
\affiliation{Quantum Simulations Group, Materials Science Division, Lawrence Livermore National Laboratory, USA}
\author{Amrith Rathnakaran}
\affiliation{Institute for Theoretical Physics, Georg-August University, Gottingen Germany}
\author{Konstantin Tamoev}
\affiliation{Center for Advanced Systems Understanding (CASUS), Helmholtz Zentrum Dresden-Rossendorf, Germany}
\author{Mani Lokamani}
\affiliation{Center for Advanced Systems Understanding (CASUS), Helmholtz Zentrum Dresden-Rossendorf, Germany}
\author{Thomas D. K\"uhne}
\affiliation{Center for Advanced Systems Understanding (CASUS), Helmholtz Zentrum Dresden-Rossendorf, Germany}
\alsoaffiliation{Institute of Artificial Intelligence, Technische Universit\"at Dresden, Germany}
\email{t.kuehne@hzdr.de}
\title{
  The CP-PAW code package for first-principles calculations from a user's perspective}
\abbreviations{IR,NMR,UV}
\keywords{American Chemical Society, \LaTeX}

\begin{document}

\maketitle
\begin{tocentry}

%Some journals require a graphical entry for the Table of Contents.
%This should be laid out ``print ready'' so that the sizing of the

%Inside the \texttt{tocentry} environment, the font used is Helvetica
%8\,pt, as required by \emph{Journal of the American Chemical

%The surrounding frame is 9\,cm by 3.5\,cm, which is the maximum
%permitted for \emph{Journal of the American Chemical Society}
%graphical table of content entries. The box will not resize if the
%content is too big: instead it will overflow the edge of the box.

%This box and the associated title will always be printed on a
%separate page at the end of the document.

\includegraphics[width=\linewidth]{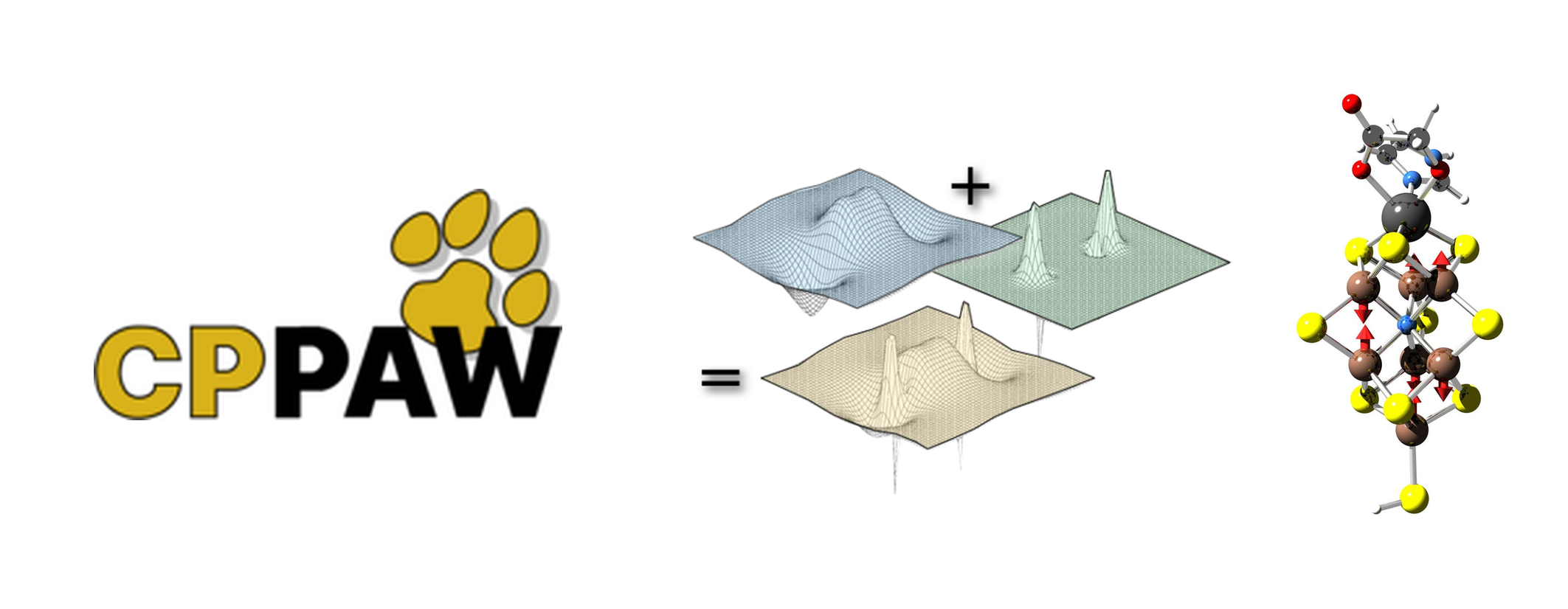}

\end{tocentry}

\begin{abstract}
  CP-PAW is a combined electronic structure and \textit{ab-initio} molecular dynamics code to perform mixed quantum and classical simulations of atomistic condensed phase systems, such as solids, liquids, and molecular systems. 
  As the name suggests, the CP-PAW code unifies the all-electron projector augmented-wave method with the Car-Parrinello approach 
  %to holistically study the electronic and nuclear structure and properties of condensed matter, as well as their dynamics. 
  to determine not only the electronic and nuclear ground state of condensed matter, but also to study their properties and dynamics. 
  In addition to briefly outlining the underlying theory, the focus will be on unique aspects of CP-PAW and how to correctly employ them as a user. 
  How to install CP-PAW using the new build system will also be briefly mentioned. 
\end{abstract}
\section{Background and Theory}

%===========================================
%\subsection{Density functional theory}
%===========================================
%\petertt{Here we need to zoom into the topic of first-principles calculations.}
%\thomastt{How about simply drop that section and directly start with PAW?}
%\petertt{Do you have something in mind? Please go ahead}

The efficient numerical representation of the Kohn-Sham wave functions and the total energy is a crucial ingredient for an efficient numerical solution of the Kohn-Sham equations, or equivalently, the direct minimization of the density functional theory (DFT) energy functional.\cite{RevModPhys.61.689, RevModPhys.71.1253, RevModPhys.87.897} In the following, we will therefore briefly review the main concepts underlying the projector augmented-wave (PAW) formalism, which is employed throughout the present work.\cite{bloechl94_prb50_17953}

%===========================================
\subsection{Projector augmented-wave method}\label{sec:PAW_abs}
%===========================================
%\konstantintt{Maybe it would be helpful to give the user an idea of some typical values for some of the variables, e.g. a plane-wave cutoff of 30Ry, or that the number of partial waves (PW) is typically set as one or two PWs per site and angular momentum or maximum angular momentum of maybe l=1 or L=2. These are some common values off the top of my head. All of which modifies the total energy functional of (18) in a way that the new E' is hopefully still sufficiently close to the exact E.}

The PAW method has been developed to make Car-Parrinello calculations work without the limitations of the pseudopotential approach, such as uncontrolled transferability and lack of information from the inner atomic region and the core electrons.\cite{vandewalle93_prb47_4244}

The augmentation of the PAW method can be expressed in terms of a linear transformation $\hat{\mathcal{T}}$, which connects physical one-particle wave functions $| \psi_n \rangle$ with numerically convenient pseudo wave functions $|\tilde{\psi}_n\rangle$, hence 
\begin{equation}
    | \psi_n \rangle = \hat{\mathcal{T}} | \tilde{\psi}_n \rangle. 
    \label{eq:paw_linear_transformation}
\end{equation}
The transformation incorporates the nodal structure into an otherwise smooth pseudo wave function. 
The latter may be expanded, as is done in the CP-PAW code, in plane waves. 

The transformation $\hat{\mathcal{T}}$ is constructed from the sum of atom-local contributions $\hat{S}_{R}$, i.e. 
\begin{equation}
    \hat{\mathcal{T}} = \hat{1} + \sum_{R} \hat{S}_{R}, 
    \label{eq:paw_transformation_operator}
\end{equation}
where the summation index $R$ is over all atoms. 

An operator such as $\hat{S}_R$ is a transformation in Hilbert space. 
It can be defined by specifying the target states for a complete set of source states.
The local contributions $\hat{S}_R$ are defined via pairs $(|\phi_\alpha\rangle,|\tilde{\phi}_\alpha\rangle)$ of partial waves that define the mapping $|\tilde{\phi}_\alpha\rangle\rightarrow |\phi_\alpha\rangle$
so that
\begin{equation}
|\phi_{\alpha}\rangle=\big(\hat{1}+\hat{S}_{R_\alpha}\big)|\tilde{\phi}_\alpha\rangle.
    \label{eq:paw_mapping}
\end{equation}
The all-electron partial wave $|\phi_\alpha\rangle$ contains the complete nodal structure of the physical wave function, while the auxiliary pseudo partial wave $|\tilde{\phi}_\alpha\rangle$ is smooth, so that it can easily be represented numerically.

Thus, the local operator $\hat{S}_R$ can be written as
\begin{equation}
\hat{S}_R=\sum_\alpha 
\Bigl(|\phi_{\alpha}\rangle -|\tilde\phi_{\alpha}\rangle\Bigr) 
\langle\tilde{p}_\alpha|
\label{eq:paw_local_contributions}
\end{equation}
with so-called projector functions $\langle\tilde{p}_\alpha|$. The latter satisfy the bi-orthogonality condition $\langle\tilde{p}_\alpha|\tilde{\phi}_\beta\rangle=\delta_{\alpha,\beta}$.

The bi-orthogonality condition ensures that the projector functions probe a certain partial wave character in a pseudo wave function: for any pseudo partial wave expansion
\begin{eqnarray}
&&|\tilde\psi\rangle=\sum_{\text{$\alpha$ with $R_\alpha=R$}} |\tilde\phi_\alpha\rangle \, c_\alpha,
\label{eq:projectionidentity}
\end{eqnarray}
the coefficients $c_\alpha$ can be recovered from $|\tilde\psi\rangle$ as scalar products $c_\alpha=\langle\tilde{p}_\alpha|\tilde{\psi}\rangle$.

The local contribution $\hat{S}_R$ is restricted to an augmentation region reaching from the nucleus to some atomic radius. 
Note, however, that the projector augmentation has, unlike other augmented-wave methods,\cite{slater37_pr51_846, korringa47_physica13_392, kohn54_pr94_1111, andersen75_prb12_3060} no clear-cut augmentation radius.

To ensure that $\hat{S}_R$ acts only in the augmentation region,
(1) the projector functions are constructed to vanish beyond the augmentation region and (2) the partial waves $(|\varphi_\alpha\rangle,|\tilde{\varphi}_\alpha\rangle)$ are pairwise identical beyond the augmentation region.

The partial wave indices $\alpha=(R_\alpha,\ell_\alpha,m_\alpha,\sigma_\alpha,n_\alpha)$ contain a site index $R_\alpha$, angular momentum quantum numbers such as $\ell_\alpha$ and $m_\alpha$, a spin index $\sigma_\alpha\in\{\uparrow,\downarrow\}$ and another index $n_\alpha$ to distinguish otherwise identical partial waves.

With so defined local contributions $\hat{S}_R$, we can execute the transformation $\hat{\mathcal{T}}$ to obtain the all-electron wave functions $|\psi_n\rangle$ from the auxiliary pseudo wave functions $|\tilde{\psi}_n\rangle$, i.e.
\begin{eqnarray}
       | \psi_n \rangle = | \tilde{\psi}_n \rangle + \sum_{\alpha} \left( | \phi_{\alpha} \rangle - | \tilde{\phi}_{\alpha} \rangle \right) \langle \tilde{p}_{\alpha} | \tilde{\psi}_n \rangle
    \label{eq:paw_projected_wave_functions}
\end{eqnarray}
for the valence states and 
\begin{equation}
    |\psi_n\rangle = | \tilde{\phi}^c_{\alpha_n}\rangle + 
    |\phi^c_{\alpha_n}\rangle - | \tilde{\phi}^c_{\alpha_n} \rangle 
    \label{eq:paw_transformed_core_states}
\end{equation}
for the core states, respectively. 
The core partial waves $|\phi^c_{\alpha}\rangle$ and $|\tilde{\phi}^c_{\alpha}\rangle$ are, like the valence partial waves, constructed for the isolated atom and they are kept frozen during the calculation. The two pseudo core wave functions $|\tilde{\phi}^c_{\alpha_n}\rangle$ in Eq.~\ref{eq:paw_transformed_core_states}, the first and last $|\tilde{\phi}^c_{\alpha_n}\rangle$, differ only in their numerical representation, namely plane waves versus atom-centered radial grids and spherical harmonics.

The wave functions are naturally divided into three components,
a plane-wave part, i.e. the pseudo wave function
$|\tilde{\psi}_n\rangle$ or the frozen pseudo core state $|\tilde{\phi}^c_{\alpha_n}\rangle$,
and two sets of one-center expansions for each atom, namely, on the one hand, the one-center expansions
\begin{eqnarray}
    |\psi^{(1)}_{R,n}\rangle=
    \sum_{\alpha; R_\alpha=R}|\phi_\alpha\rangle
    \langle\tilde{p}_\alpha|\tilde\psi_n\rangle
\end{eqnarray}
of the valence states and the frozen core states $|\phi_{\alpha_n}\rangle$, as well as, on the other hand, the one-center expansions 
\begin{eqnarray}
    |\tilde{\psi}^{(1)}_{R,n}\rangle=
    \sum_{\alpha; R_\alpha=R}|\tilde{\phi}_\alpha\rangle
    \langle\tilde{p}_\alpha|\tilde\psi_n\rangle
\end{eqnarray}
of the valence pseudo states and the frozen pseudo core states $|\tilde{\phi}_{\alpha_n}\rangle$.
The one-center expansions are represented by functions on a radial grid multiplied by spherical harmonics.

%===========================================
\subsubsection{Basisstates}
%=============================================
The basis functions of the CP-PAW code are projector-augmented plane waves, in addition to the frozen core states of the isolated atoms.
A projector-augmented plane-wave $W_{\vec{G},\sigma}(\vec{r},\sigma')$
with wave vector $\vec{G}$ and spin $\sigma\in\{\uparrow,\downarrow\}$ has the form
\begin{eqnarray}
W_{\vec{G},\sigma}(\vec{r},\sigma')
&=&\e{i\vec{G}\vec{r}}\delta_{\sigma,\sigma'}
\nonumber\\
&+&
\sum_\alpha \Bigl(
\phi_\alpha(\vec{r},\sigma')-\tilde{\phi}_\alpha(\vec{r},\sigma')\Bigr)
\nonumber\\
&&\times
\tilde{p}^*_\alpha(\vec{G},\sigma),
\end{eqnarray}
where the Fourier coefficients of the projector functions are 
\begin{eqnarray}
\tilde{p}_\alpha(\vec{G},\sigma)
% &=&
% \langle\tilde{p}_\alpha|\vec{G},\sigma\rangle
% \frac{1}{\langle\vec{G},\sigma|\vec{G},\sigma\rangle}
% \nonumber\\
&=&
\underbrace{
\frac{1}{V}\int_V d^3r\; \tilde{p}_\alpha(\vec{r}-\vec{R}_\alpha
,\sigma)
\e{-i\vec{G}(\vec{r}-\vec{R}_\alpha)}
}_{\text{form factor}}
\nonumber\\
&&\times\underbrace{
\e{-i\vec{G}\vec{R}_\alpha}
}_{\text{struct. fact.}}.
\end{eqnarray}
The wave vector $\vec{G}$ is the sum of a k-point and a reciprocal lattice vector.
Together with the atomic core states $\phi^c_\alpha(\vec{r},\sigma')$, the augmented plane waves up to a specified cutoff span the Hilbert space explored during the optimization or dynamics, respectively.

%===============================================
\subsubsection{Natural orbitals versus energy eigenstates and the frozen-core approximation}
%===============================================
Let us dwell a little on the distinction between natural orbitals of the Kohn-Sham system, on the one hand, and the eigenstates of the Kohn-Sham Hamiltonian, on the other.
This distinction is particularly relevant and a common source of misunderstandings in the context of the Car-Parrinello method in general and the frozen-core approximation in particular.

The energy of DFT is usually expressed as a functional of the one-particle reduced density matrix $\hat{\rho}^{(1)}$ of the Kohn-Sham system, which, in turn, is represented by its natural orbitals $|\psi_n\rangle$ and occupations $f_n$,\cite{loewdin55_pr97_1474,davidson72_rmp44_451} i.e.
\begin{eqnarray}
\hat{\rho}^{(1)}=\sum_n |\psi_n\rangle f_n\langle\psi_n|,
\label{1-RDM}
\end{eqnarray}
where $\langle\psi_m|\psi_n\rangle=\delta_{m,n}$.

The minimum principle of DFT determines the natural orbitals only up to gauge transformations, which leave the one-particle reduced density matrix, and thus the energy, invariant. 
These are unitary transformations between states with the same occupation.

% After optimization, the wave functions described in Eq.~\ref{eq:paw_projected_wave_functions} and Eq.~\ref{eq:paw_transformed_core_states} are not necessarily eigenstates of the Kohn-Sham Hamiltonian, but eigenstates of the one-particle reduced density matrix, so-called natural orbitals (of the Kohn-Sham system).\cite{loewdin55_pr97_1474,davidson72_rmp44_451}There is an intrinsic ambiguity in their definition with respect to so-called gauge transformations, which leave the one-particle reduced density matrix invariant.

The eigenstates $|\psi^{eig}_n\rangle$ of the Kohn-Sham Hamiltonian $\hat{h}_{\text{eff}}$ can be obtained from the natural orbitals after diagonalizing the Hamiltonian, so that
%\begin{eqnarray}
%|\psi^{eig}_n\rangle&=&\sum_{m}|\psi_m\rangle U_{m,n}
%\nonumber\\
%\text{with}&&\sum_{j}
%\langle\psi_m|\hat{h}_{\text{eff}}|\psi_j\rangle U_{j,n}
%=U_{m,n}\epsilon_n.
%\end{eqnarray}
\begin{subequations}
\begin{equation}
|\psi^{eig}_n\rangle = \sum_{m}|\psi_m\rangle U_{m,n}
\end{equation}
with $U_{m,n}$ defined by the diagonalization
\begin{equation}
\sum_{j} \langle\psi_m|\hat{h}_{\text{eff}}|\psi_j\rangle U_{j,n}
=U_{m,n}\epsilon_n,
\end{equation}
\end{subequations}
where $\epsilon_n$ are the Kohn-Sham energies.

As will be seen in the following, the Car-Parrinello method determines the ground state strictly by minimizing the total energy functional. 
The resulting wave functions are natural orbitals, but not necessarily eigenstates of the Kohn-Sham Hamiltonian.
If desired, the diagonalization is done as post-processing.

The distinction between natural orbitals and eigenstates of the Kohn-Sham Hamiltonian is particularly important in the context of the frozen-core approximation. 
The frozen-core approximation uses the atomic core states as \textit{basis functions} for the wave function in the molecule or solid.
It does not, however, claim that these states are \textit{eigenstates} of the Kohn-Sham Hamiltonian.
A comparison of frozen atomic core states with core wave functions is therefore invalid, because the latter are, even in the frozen core approximation, a superposition of all core and fully occupied valence states.
This superposition is rarely done, on purpose, in order to avoid the need to discuss the role of core states in chemical bonding.

The quality of the frozen-core approximation depends on the quality of the atomic core wave functions used. 
In the CP-PAW code, these states are numerically exact, so that the errors are of second order in the deviation from the isolated atom.

Nevertheless, the PAW method is not limited to the frozen-core approximation, as demonstrated by Marsman and coworkers.\cite{marsman06_jcp125_104101}

%=================================================
\subsubsection{Expectation values and total energy}
%=================================================
After having gone into some detail regarding the basis set, we will be brief with respect to its use in evaluating expectation values and the energy functional.
It shall suffice to say that the division of the wave function into plane-wave parts and one-center expansions can be carried out down to expectation values of one-particle operators $\hat{A}$ via
\begin{eqnarray}
    \langle\psi|\hat{A}|\psi\rangle
    &=&\langle\tilde{\psi}|\hat{A}|\tilde{\psi}\rangle
    \nonumber\\
    &&\hspace{-2cm}+\sum_R
    \Bigl(\langle\psi^{(1)}_R|\hat{A}|\psi^{(1)}_R\rangle
    -\langle\tilde{\psi}^{(1)}_R|\hat{A}|\tilde{\psi}^{(1)}_R\rangle\Bigr)
\end{eqnarray}
and the total energy
\begin{eqnarray}
    E[\{|\psi_n\rangle,f_n\}]
    &=&
   \tilde{E}[\{|\tilde\psi_n\rangle,f_n\}]
   \nonumber\\
   &&\hspace{-3cm}+\sum_R \Bigl(E^{(1)}_R[\{|\psi^{(1)}_{R,n}\rangle,f_n\}]
   -\tilde{E}^{(1)}_R[\{|\tilde\psi^{(1)}_{R,n}\rangle,f_n\}]\Bigr).
   \nonumber\\
\end{eqnarray}
These expectation values and the total energy include both, the valence and core states, represented as in Eq.~\ref{eq:paw_projected_wave_functions} and Eq.~\ref{eq:paw_transformed_core_states}, respectively.

For the expectation values, we exploit that (1) the all-electron and pseudo partial waves are, per construction, pairwise identical beyond the augmentation region $\Omega_R$ centered at site $R$, so that $\psi^{(1)}_{R,n}(\vec{r},\sigma)-\tilde\psi^{(1)}_{R,n}(\vec{r},\sigma)=0$ for $\vec{r}\notin\Omega_R$, and that (2) the one-center expansion of the pseudo wave function is accurate in the augmentation region $\Omega_R$, i.e.   $\tilde{\psi}_n(\vec{r})-\tilde{\psi}^{(1)}_{R,n}(\vec{r})\approx0$ for $\vec{r}\in\Omega_R$.

Dividing the total energy into three contributions proceeds analogously. The main difficulty is the long-ranged Coulomb interaction. 
It is tackled by constructing a so-called compensation density, which is added to the pseudo density and its one-center expansion. 
This decouples the one-center expansions electrostatically, and transfers this term into the plane-wave part, where it is combined with the Hartree term.
For further details on how to represent the energy and expectation values, we refer to the original publication.\cite{bloechl94_prb50_17953}

%=================================================
%\subsubsection{Approximations of the PAW method}
%=================================================
%All approximations, even numerical ones, are defined on the level of the energy, respectively the level of the Lagrangian. 
%Forces and potentials are derived analytically from such an approximate energy functional.
%\begin{itemize}
%    \item a cutoff for the plane-wave part of the augmented plane waves
%    \item a set of partial waves and projector functions that define the augmentation
%    \item atomic core states as basis set for the core wave functions.
%\end{itemize}
%\thomastt{Is this section necessary?}

%===============================================
\subsubsection{Setup construction of the augmentation}
%===============================================
The augmentation is defined for each atom type by
\begin{itemize}
\item a set of partial wave pairs $|\phi_\alpha\rangle$ and $|\tilde{\phi}_\alpha\rangle$, and the corresponding projector functions $\langle\tilde{p}_\alpha|$,
\item the density of the frozen core states and its pseudized version,
\item {$V_{add}$}, a potential limited to the augmentation regions, which acts on the pseudo density and its one-center expansion,
\item {the extent of the compensation densities}, used to transfer the intersite Coulomb interaction between the one-center densities into the plane-wave part, and
\item the atomic number.
\end{itemize}

The CP-PAW code constructs the augmentation on-the-fly during the initialization phase of each calculation based on a small set of parameters set by the user.
Hence, no external datasets are required.
This ensures consistency with internal parameters of the calculation and furthermore gives the user control over the augmentation.

An example for silicon is given below

{\scriptsize
\begin{verbatim}
!SPECIES   NAME='SI'  NPRO=2 2 1 LRHOX=4 RAD/RCOV=1.4
  !AUGMENT ID='MY_SI' EL='SI' CORE=''
           TYPE='NDLSS' RBOX/RCOV=1.2 RCSM/RCOV=.25
           RCL/RCOV=0.75 0.75 0.75 0.75
    !GRID  DMIN=1.E-6 DMAX=.15 RMAX=9. !END
    !POT   POW=3. RC/RCOV=0.75  !END
    !CORE  POW=2. RC/RCOV=0.75 !END
  !END
!END
\end{verbatim}
}

Let us only mention the most important parameters: \verb|NPRO=2 2 1| specifies the number of partial wave pairs that shall be used per angular momentum for $\ell=0,1,2$. The parameters \verb|RCL/RCOV=0.75 0.75 0.75 0.75| specify the matching radii of the partial wave pairs in units of the covalent radius $r_{cov}$ of that atom. The parameter \verb|CORE=''| allows defining the division into core- and valence-electron shells. 

For each atom type, a self-consistent atom calculation is performed, which serves the construction of partial waves, projector functions, etc. 
In order to avoid breaking the symmetry through the augmentation, this atom is completely spherical and does not exhibit any spin polarization.
A set of numerically exact solutions of the Schr\"odinger equation (respectively, the Dirac equation) is collected as all-electron partial waves. 
The corresponding pseudo partial waves are then chosen according to some recipe, which removes the oscillatory behavior from the physical wave functions, while maintaining the same functional form beyond the augmentation region.
This recipe determines the well-behavedness of the pseudo partial waves. 

The projector functions are constructed from a set of smooth functions $\langle{f}_\alpha|$, which are smooth and localized in the augmentation region. 
The form 
\begin{eqnarray}
    \langle\tilde{p}_\alpha|=\sum_\beta A_{\alpha,\beta}\langle{f}_\beta|
\end{eqnarray}
with $\sum_{\gamma}A_{\alpha,\gamma}\langle{f}_\gamma|\tilde\phi_\beta\rangle=\delta_{\alpha,\beta}$ enforces the bi-orthogonality condition. The freedom of choice for the functions $\langle{f}_\alpha|$ is used to ensure that the PAW method satisfies the Schr\"odinger equation for the set of partial waves exactly. In the non-relativistic case, this closure condition implies
\begin{eqnarray}
  |f_{\alpha}\rangle=\Bigl(\frac{\hat{\vec{p}}^2}{2m_e}+\hat{\tilde{v}}-\epsilon^{at}_\alpha\Bigr)|\tilde\phi_\alpha\rangle,
\end{eqnarray}
where $\tilde{v}$ is an auxiliary local potential, which is described later, and $\epsilon_\alpha^{at}$ is the atomic energy of the all-electron partial wave.

Partial waves and projector functions are expressed in terms of radial functions $g_\alpha(r)$ on atom-centered radial grids, multiplied by real spherical harmonics, such as
\begin{eqnarray}
    \langle\vec{r},\sigma|\phi_{\alpha}\rangle
    =g_\alpha(|\vec{r}-\vec{R}_\alpha|)
    Y_{\ell_\alpha,m_\alpha}(\vec{r}-\vec{R}_\alpha)\delta_{\sigma,\sigma_\alpha}.
\end{eqnarray}
The radial grid is a shifted logarithmic grid $r(i)=r_1(\e{\gamma(i-1)}-1)$ with $i\in\lbrace 1,2,\ldots,N\rbrace$, which is sufficiently fine to resolve the shape of the wave function inside the nucleus. Bessel transforms map projector functions from a radial grid in real space onto a radial grid in reciprocal space.

The construction of setups can lead to instabilities. One common cause is the overcompleteness problem. 
If the number of projector functions exceeds the number of degrees of freedom of the pseudo wave function in the augmentation region, some partial waves have undetermined prefactors. 

The second problem is related to so-called ghost states, which are a common problem for augmented wave methods and fully non-local pseudopotentials.
The construction of the augmentation, aka the setup construction, needs to take care to avoid these problems.

%==========================================================================
\subsection{Car-Parrinello method}
%==========================================================================
The CP-PAW code differs from many other electronic-structure program packages by using the fictitious Lagrangian approach towards \textit{ab-initio} molecular dynamics.\cite{car85_prl55_2471, kuhne2014second} The first two letters of the name of the CP-PAW code are a tribute to the inventors of this technique, Roberto Car and Michele Parrinello.
While the Car-Parrinello method requires a somewhat different mindset compared to other techniques, 
its operations are fairly robust and physically transparent. And although coding requires special precautions and care, it leads to a rather simple and modular code structure.

The Car-Parrinello method is based on the simultaneous propagation of nuclei and electronic wave functions. The nuclei are classical particles, whereas the electronic one-particle wave functions are considered as time-dependent classical fields. 
Instead of propagating the wave functions with the time-dependent Schr\"odinger equation, the Car-Parrinello method uses a second-order differential equation like the Newtonian dynamics of the nuclei.\cite{kuhne2007efficient, kuhne2018disordered} This coupled electron-ion dynamics ensures that the wave functions can be kept close to the instantaneous (thermal) ground state as required by DFT. 

The corresponding equations of motion for the nuclei and electronic single-particle wave functions read as
\begin{eqnarray}
    M_R\ddot{\vec{R}}_R&=&\vec{F}_R
    \nonumber\\
    \hat{m}_\psi |\ddot{\tilde{\psi}}_n\rangle&=&
    -\hat{\tilde{h}}_{eff}|\tilde{\psi}_n\rangle
    \nonumber\\
    &+&\sum_m\hat{\tilde{O}}|\tilde{\psi}_m\rangle \Lambda_{m,n}\frac{1}{f_n}, 
    \label{eq:eqmcp}
\end{eqnarray}
where the atomic positions $\vec{R}(t)$ and the Kohn-Sham pseudo wave functions $|\tilde\psi_n(t)\rangle$ are determined dynamically as a function of time $t$. Moreover, $\hat{\tilde{O}}$ is an overlap operator, which arises as a consequence of the augmentation.
 
The occupations of the Kohn-Sham wave functions are denoted as $f_n$. At variance to common practice, the occupations also enter the fictitious kinetic energy of the wave functions.
The atomic masses are denoted as $M_R$, while $\hat{m}_\psi$ is a fictitious mass operator for the wave function dynamics.

The nuclear forces $\vec{F}_R$ and the Kohn-Sham Hamiltonian, respectively  $\hat{\tilde{h}}_{\text{eff}}|\tilde\psi_n\rangle$, are obtained from the partial derivatives of the energy functional of DFT with respect to atomic positions and the Kohn-Sham wave functions. 
The conceptually most complex aspect are the Lagrange multipliers $\Lambda_{m,n}$, which enforce the orthonormality condition of the Kohn-Sham wave functions during the dynamics.

In the Car-Parrinello method, all dynamical equations of motion are deduced rigorously from a single action functional of the trajectories of all dynamical variables, i.e.
\begin{eqnarray}
S=\int dt\; \mathcal{L}\big(
\vec{R},\dot{\vec{R}},|\tilde{\psi}_n\rangle,|\dot{\tilde{\psi}}_n\rangle, ...,t
\big).
\label{eq:actionasfuncoflagrfunc}
\end{eqnarray}

The Lagrange function $\mathcal{L}$ has the form
\begin{eqnarray}
    \mathcal{L}&=&\sum_n f_n\langle\dot{\tilde{\psi}}_n|\hat{m}_\psi|\dot{\tilde{\psi}}_n\rangle
    +\sum_R \frac{1}{2}M_R \dot{\vec{R}}^2
    \nonumber\\
&-&E_{DFT}[R,\hat{\mathcal{T}}|\tilde{\psi}_n\rangle,f_n]
\nonumber\\
&+&\sum_{m,n}\Lambda_{m,n}\Bigl(\langle\tilde{\psi}_n|\hat{\tilde{O}}(\vec{R})|\tilde{\psi}_m\rangle-\delta_{m,n}\Bigr).
\end{eqnarray}

The requirement that the action is stationary with respect to variations of the physical trajectory translates into the Euler-Lagrange equations, which are the coupled electron-ion equations of motion shown in Eq.~\ref{eq:eqmcp} with nuclear forces 
%The equations of motion shown in Eq.~\ref{eq:eqmcp} are obtained from the requirement that the action is stationary with respect to variations of the physical trajectory. This translates into Euler-Lagrange equations. The latter are identical to Eq.~\ref{eq:eqmcp} with forces
%This translates into Euler-Lagrange equations, eventually leading to the nuclear forces ...
\begin{eqnarray}
\vec{F}_R=-\vec{\nabla}_R E_{DFT}+\sum_{m,n}\Lambda_{m,n}
\langle\tilde{\psi}_n|\vec{\nabla}_R \hat{\tilde{O}}|\tilde{\psi}_m\rangle
\end{eqnarray}
and an (pseudo) effective Hamiltonian that satisfies
\begin{eqnarray}
\hat{\tilde{h}}_{\text{eff}}|\tilde{\psi}_n\rangle f_n
=\frac{\partial E_{DFT}[R,\hat{T}|\tilde{\psi}_n\rangle,f_n]}{\partial\langle\tilde{\psi}_n|}.
\end{eqnarray}

The action principle provides an elegant way to integrate different systems into one consistent scheme.
Such additional degrees of freedom may be the electronic occupations, unit-cell parameters, thermo- and barostat variables, or even additionally coupled physical systems as in mixed quantum-classical mechanics methods.
Friction terms can be added, which turns the energy-conserving dynamics into an optimization scheme that can be considered as dynamical simulated annealing.\cite{kirkpatrick1983optimization, heller2008inflationary}

%=========================================================
\subsubsection{Verlet algorithm}
%=========================================================
The equations of motion represented in Eq.~\ref{eq:eqmcp} are solved using the Verlet algorithm, which combines numerical stability and computational efficiency with exact time-reversibility and preservation of the symplectic form in phase space.\cite{verlet67_pr159_98}
The latter is important to maintain global energy conservation.

The Verlet algorithm replaces the first and second time derivatives by the time-symmetric differential quotients. 
For the damped equation of motion for a vector $\vec{x}$ of dynamical variables
\begin{eqnarray}
\mathbf{m}\ddot{\vec{x}}&=&\vec{F}-\mathbf{m}\gamma\dot{\vec{x}}\;
\end{eqnarray}
with a mass tensor $\mathbf{m}$ and a friction parameter $\gamma$,
this yields
\begin{eqnarray}
\vec{x}(t+\Delta)&=&\vec{x}(t)\frac{2}{1+a}
-\vec{x}(t-\Delta)\frac{1-a}{1+a}
\nonumber\\
&&+\mathbf{m}^{-1}\vec{F}(t)\frac{\Delta^2}{1+a},
\end{eqnarray}
where $\Delta$ is the discretized time step and $a=\gamma\Delta/2$ encodes the friction.
With $\vec{x}(t)$, we denote the trajectory of the dynamical degrees of freedom, which may be atomic positions, the electronic wave function, etc.
 The friction parameter $\gamma$ is chosen individually for each type of degree of freedom.

When the friction parameter is zero $(a=0)$, the system will undergo an energy-conserving dynamics, whereas
the choice $a=1$ leads to a steepest-descent dynamics corresponding to an infinite friction parameter $\mathbf{m}\gamma$. 
A positive value of $a$ implies damping the system, while a negative parameter $a$ will heat the system. 

\begin{figure}[!hbt]
\includegraphics[width=\linewidth]{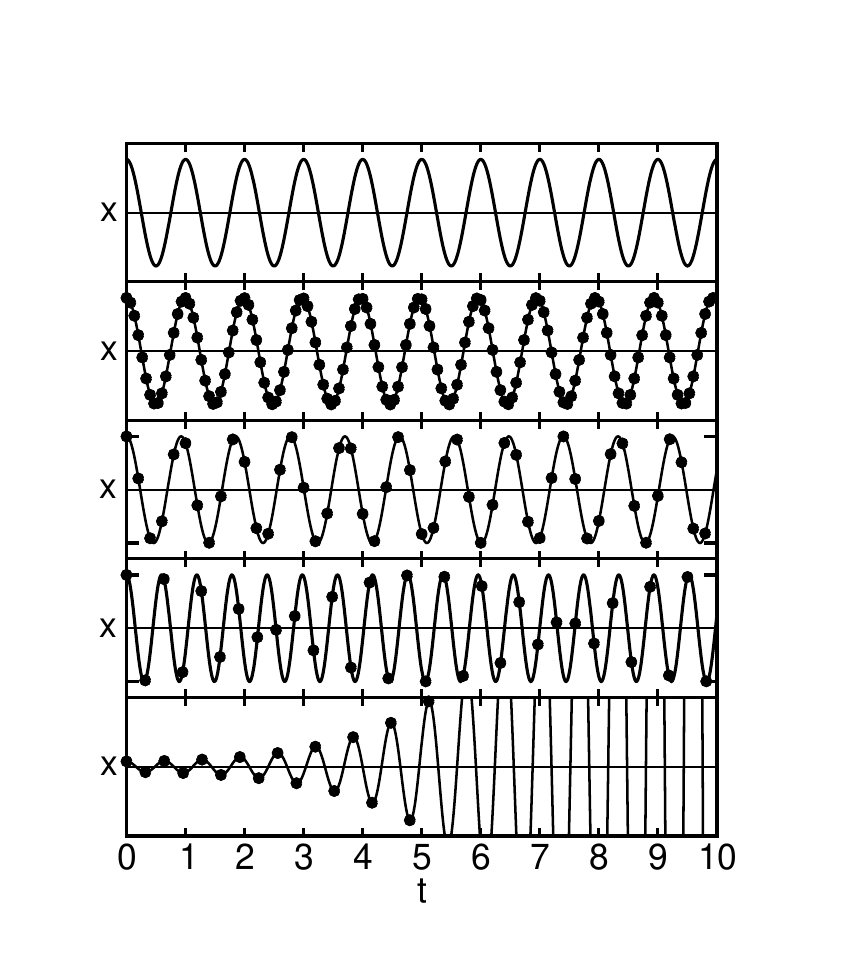}
\caption{\label{fig:verlettra} Trajectory $x(t)$ of a one-dimensional harmonic oscillator obtained with the Verlet algorithm for different time steps $\Delta$ with $\Delta/T_0=0,\frac{1}{15},\frac{1}{5},0.99/\pi,1.01/\pi$ from top to bottom. $T_0$ is the exact oscillation period. The dots represent the calculated points, while the line is a plane-wave passing through the calculated points. The top row is the exact result. In the bottom row, the time step exceeds the stability limit, resulting in an exponential divergence. Within the stability limit, i.e. rows 2,3 and 4, the trajectory maintains the qualitative behavior with a frequency increasing with the time step.}
\end{figure}

The most important aspects of the dynamics can already be learned from a one-dimensional harmonic oscillator. 
The harmonic oscillator allows for the continuous and discretized equations of motion to be solved analytically. 
The trajectory obtained with increasing time steps is shown in Fig.~\ref{fig:verlettra}. 
The trajectory remains sinusoidal up to the stability limit, beyond which the trajectory blows up exponentially.

\begin{figure}[!hbt]
\includegraphics[width=\linewidth]
{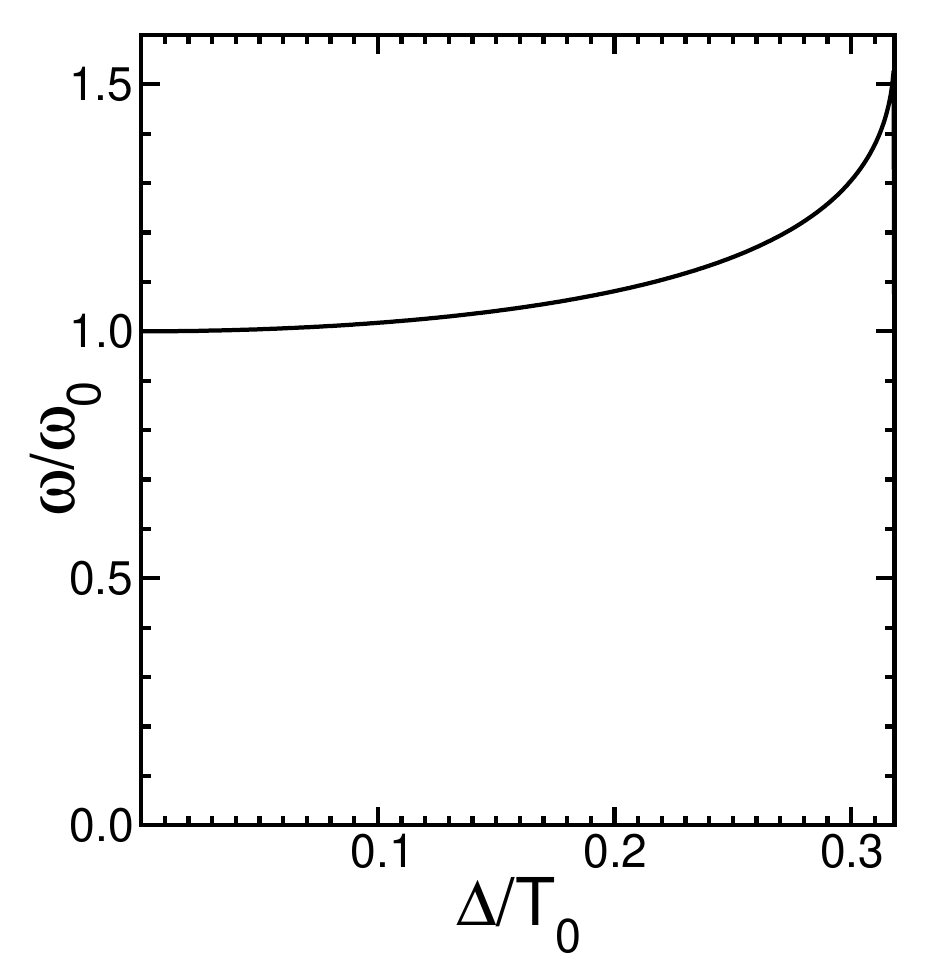}
\caption{\label{fig:stb1}Frequency of the discretized one-dimensional harmonic oscillator, as a function of the time step $\Delta$
in terms of the exact oscillation period $T_0=2\pi/\omega_0$.}
\end{figure}

As shown in Fig.~\ref{fig:stb1}, the frequency of the discretized dynamics increases with increasing time step. 
With only about ten time steps per period, the frequency of the discretized dynamics is accurate within two percent.

Most importantly, the trajectory becomes unstable for time steps beyond $\Delta>2/\omega_0$, where $\omega_0$ is the circular frequency of the harmonic oscillator. Hence, the Verlet algorithm has a stability limit at 
\begin{eqnarray}
 \Delta_{stab}=\frac{2}{\omega_0}.
\end{eqnarray}
This implies that a stable trajectory is obtained already with little more than three time steps per oscillation period.

%================================================================
\subsubsection{Optimization}
\label{sec:optimization}
%================================================================
The CP-PAW code uses the very same dynamical scheme, namely damped dynamics, to optimize the electronic and nuclear degrees of freedom.
Increasing the friction is the obvious recipe to reach rapid convergence towards the minimum. 
However, this is only so until one reaches critical damping, which separates damped oscillations from the pure exponential decay of an overdamped relaxation. 
The convergence in the overdamped regime is disappointingly slow. 
Hence, an efficient optimization will choose a friction parameter below the optimum friction. 

The optimum friction value, resulting in critical damping is 
\begin{eqnarray}
a_{opt}=\omega\Delta,
\label{eq:aopt}
\end{eqnarray}
 where the friction value is defined as $a=\gamma\Delta/2$.

\begin{figure}[!hbt]
\begin{center}
\includegraphics[width=\linewidth]{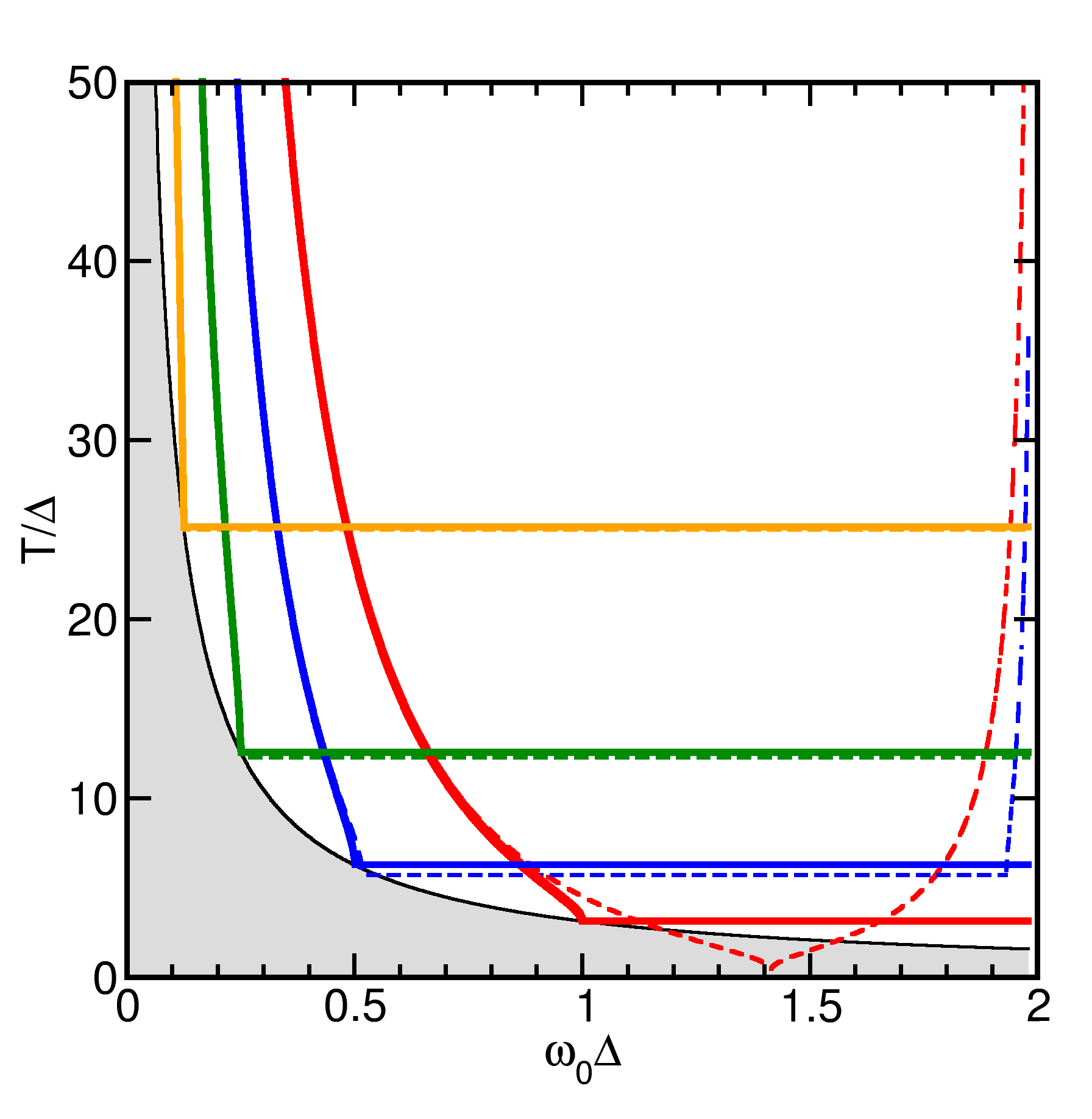}
\end{center}
\caption{\label{fig:decaytimevsomega}Decay time $T$ of the total
  energy as a function of vibrational frequency $\omega_0$. The decay
  time describes the convergence of the total energy as
  $E(t)\sim\exp(-\frac{t}{T})$. The friction parameters
  $a=\alpha\Delta/2$ are red for $a=1$, blue for $a=0.5$, green for $a=0.25$ and
  orange for $a=0.125$, respectively. The stability limit of the
  Verlet algorithm is $\omega_0\Delta=2$, whereas the grey region is not accessible with any
  friction parameter.
  The dashed lines are the results from the
  discretized equations, while the full lines are from the continuous
open  trajectories.}
\end{figure}

For systems with a wide frequency spectrum, the optimum friction value separates overdamped low-frequency modes from the higher-frequency ones that undergo damped oscillations. 
The energies of the high-frequency modes decrease exponentially with time, proportional to $\exp(-2\gamma t)$. 
In contrast, the convergence rate of the low-frequency modes is so low that they are effectively frozen in. This behavior is demonstrated in Fig.~\ref{fig:decaytimevsomega}, which shows the decay time $T=1/(2\gamma)$ as a function of the mode frequencies for selected friction values $a$. 

This suggests the optimization strategy implemented in the CP-PAW code: starting with a high friction, the friction is scaled down over time in order to also converge modes with lower frequency. 
By lowering the friction parameter, the frequency window of modes, which are damped out effectively, is increased.
The strategy of continuous scaling of the friction parameter is augmented by two techniques:
To avoid wasting time, a second, higher friction parameter takes effect whenever the potential energy increases in time.
Finally, a minimum friction value can be set, which has proven useful in particular for the wave function dynamics.

%================================================
\subsubsection{G-dependent mass}
%================================================
The dynamics of the wave functions can be studied for a model, namely a free electron gas with an effective Hamiltonian
\begin{eqnarray}
    \hat{h}=\sum_{\vec{G}}|\vec{G}\rangle
    \left(V_0+\frac{\hbar^2\vec{G}^2}{2m_e}\right)
    \langle\vec{G}|.
\end{eqnarray}
An effective mass
\begin{eqnarray}
    \hat{m}_\psi=\sum_{\vec{G}}|\vec{G}\rangle
    \frac{1}{\omega^2}
    \Bigl(V_0+\frac{\hbar^2\vec{G}^2}{2m_e}\Bigr)
    \langle\vec{G}|
\end{eqnarray}
will result in a dynamics for which the wave function coefficients oscillate with a target frequency $\omega$.\cite{tassone94_prb50_10561} 
The target frequency can be adapted so that the oscillation period extends over a fixed number $N$ of time steps, i.e.
$\omega={2\pi}/({N\Delta})$. 
Given the assumptions of the model, a period with 10 time steps appears to be a reasonable choice.

In CP-PAW, the G-dependent mass is extracted from the pseudo wave functions and is suggested as the default value for the G-dependent mass enhancement.

%==================================================================
\subsubsection{Mass renormalization}
\label{sec:massrenormalization}
%==================================================================
In the fictitious-Lagrangian formalism, an atom should be considered as a quasi-particle consisting of a nucleus and the surrounding wave function cloud. %\cite{car85_prl55_2471}
The wave functions attached to the nuclei add to the effective mass of the atoms, making them heavier than the bare nuclei.

The correct dynamics is obtained only after the bare masses of the nuclei are renormalized so that the net mass of nuclei including that of the attached wave functions equals the physical mass.\cite{bloechl94_prb50_17953,bloechl02_prb65_104303}
This effective mass of the wave function cloud is calculated for a rigid atom model\cite{bloechl92_prb45_9413,bloechl94_prb50_17953,bloechl02_prb65_104303} and removed from the nuclear mass, i.e. 
\begin{equation}
  \delta M = \frac{2}{3\hbar^2} \sum_n f_n\langle{\tilde{\psi}}_n|
 \hat{\vec{p}}\,\hat{m}_\psi\hat{\vec{p}}\,
  |{\tilde{\psi}}_n\rangle.
\end{equation}
Without this correction, the vibrational frequencies in dynamical simulations are systematically too low. The impact of mass renormalization on the nuclear forces has been investigated also by Tangney.\cite{tangney02_jcp116_14, tangney2006theory}

%==================================================================
\subsubsection{Thermostatting the atoms}
%==================================================================
In a canonical ensemble, the system undergoes energy fluctuations of a well-defined magnitude. In his seminal work, Shūichi Nos\'e showed how the canonical ensemble can be established rigorously in a molecular dynamics simulation through what is nowadays called the Nos\'e thermostat.\cite{nose84_jcp81_511,nose84_molphys52_255,nose91_progtheorphyssupp103_1}  
The Nos\'e thermostat adds a single dynamical friction variable $x(t)$ that 
acts like a heat bath and, thus, drives the energy fluctuations of the canonical ensemble.
Hoover translated the original formulation of Nos\'e, which used a scaled time variable, into real time.\cite{hoover85_pra31_1695}

This Nos\'e-Hoover thermostat can be cast in the form\cite{bloechl92_prb45_9413}
\begin{eqnarray}
\mathbf{M}\ddot{\vec{R}}&=&\vec{F}-\mathbf{M}\dot{\vec{R}} \dot{x}
\nonumber\\
Q\ddot{x}&=&2\biggl(\frac{1}{2}\dot{\vec{R}}\mathbf{M}\dot{\vec{R}}-\frac{1}{2}gk_BT\Biggr).
\end{eqnarray}

The Nos\'e-Hoover thermostat has a characteristic time scale for the energy fluctuations, namely $2\pi/\omega$ with
\begin{eqnarray}
\omega=\sqrt{\frac{2gk_BT}{Q}}.
\end{eqnarray}
On the one hand, this time scale must be several times shorter than the eventual simulation time, so that the thermal fluctuations are properly averaged. 
On the other hand, however, this time scale should be large compared to that of the physical processes of interest. 

Thermostats rely on a thermal coupling between the parts of the system. 
Without thermal coupling, the thermostat maintains an average temperature by heating up one subsystem above the target temperature, while cooling another subsystem below the target.
As a consequence, thermal equilibrium will never be reached.
Thus, caution is required when thermostatting weakly coupled subsystems.

A particularly serious case is the so-called flying ice-cube effect.\cite{harvey1998flying, braun2018anomalous}
Therein, global translations of solids, and for isolated molecules also the rotational degrees of freedom, are decoupled from the remaining system, which in turn prevents thermalization.
The remedy is to suppress these uncoupled degrees of freedom using constraints on a global translation, and, for molecules, also on the global rotation.

%===============================================================
\subsubsection{Wave function thermostat}
%===============================================================
A Car-Parrinello \textit{ab-initio} molecular dynamics simulation corresponds to a non-equilibrium system with "hot" atoms and "cold" wave functions.
In order to make such a simulation stationary, we combine a Nos\'e-Hoover thermostat controlling the temperature of the atoms with another one keeping the wave functions close to the ground state, so that the requirements of DFT are satisfied.
\cite{bloechl92_prb45_9413,bloechl02_prb65_104303}

While the atoms are moving in a molecular dynamics simulation, the wave functions have a certain fictitious kinetic energy when they follow the atomic motion adiabatically.
The attempt to remove this kinetic energy by a friction acting on the wave functions behaves like a strong friction acting indirectly on the atoms.
It is the additional kinetic energy beyond this adiabatic motion, which corresponds to thermal fluctuations and which needs to be damped out in Car-Parrinello \textit{ab-initio} molecular dynamics simulations.

The original two-thermostat formulation\cite{bloechl92_prb45_9413} has been developed further.\cite{bloechl02_prb65_104303}
The purpose of the wave function thermostat is not thermalization, but the transfer of thermal energy from the wave functions into the atomic motion. 
The new wave function thermostat inherits much of its form from the Nos\'e-Hoover thermostat, but it is no longer a thermostat:
\begin{itemize}
\item It never adds energy to the wave function dynamics
\item it does not add and remove energy to and from the system, but it transfers energy from the wave function degrees of freedom to the nuclear degrees of freedom.
\end{itemize}

The equations of motion with the two-thermostat formulation and mass renormalization become somewhat more involved \cite{bloechl02_prb65_104303}
\begin{subequations}
\begin{eqnarray}
&&  \hat{m}_\psi\vert\ddot{\tilde\psi}_n\rangle 
   = - \hat{\tilde{h}}_{eff} \vert\tilde\psi_n\rangle
   + \sum_{m} \tilde{O}|\tilde\psi_m\rangle \Lambda_{m,n}\frac{1}{f_n}
\nonumber\\
&&
\hspace{15mm}- \hat{m}_\psi\vert\dot{\tilde\psi}_n
   \rangle \dot x_\Psi 
\\ \nonumber \\
&& (M_i-K_{i,i}) \ddot R_i = F_i 
\nonumber\\
 &&\hspace{15mm}- M_i\dot{R}_i\dot{x}_R  + K_{i,i}\dot{R}_i\dot{x}_\psi
% \nonumber\\&&
% +(\sum_{j}
% \frac{\partial K_{i,i}}{\partial R_j} \dot{R_j}
% \dot{R_i}
% -\frac{1}{2}\sum_{j}\frac{\partial K_{j,j}}{\partial R_i}\dot{R_j}^2)
\\ \nonumber \\
  &&Q_\psi \ddot x_\psi
  = 2 \theta(\dot{x}_\Psi)
  \nonumber\\
  &&\hspace{3mm}\times
\Bigl[ \sum_{n} f_n\langle\dot{\tilde\psi}_n |m_\psi| \dot{\tilde\psi}_n \rangle 
  - \sum_{i}\frac{1}{2}K_{i,i}\dot{R}_i^2 \Bigr]
\\ \nonumber \\
 &&  Q_R\ddot x_R
   = 2 \Bigl(\sum_{i}\frac{1}{2}M_i\dot R_i^2-\frac{1}{2}gk_BT\Bigr),
\end{eqnarray}
\end{subequations}
where $\theta(x)$ is the Heaviside step function.
The effective mass of the wave function cloud $K_{i,i}$ is calculated beforehand from the atomic pseudo wave functions as %\petertt{Eq.19 of \cite{bloechl02_prb65_104303}}
\begin{eqnarray}
    K_{i,i}=\frac{2}{3}\sum_n f_n\int_0^\infty\hspace{-3mm} dG\;|\tilde\psi_n(G)|^2 m_\psi^0(G^2+cG^4),
\end{eqnarray}
where the G-dependent wave function mass is %\petertt{Eq.16 of \cite{bloechl02_prb65_104303}}
\begin{eqnarray}
\hat{m}_\psi(\vec{G})=\sum_{\vec{G}}
|\vec{G}\rangle m_\psi^0\left(1+c|\vec{G}|^2\right) \langle\vec{G}|.
\end{eqnarray}

As discussed earlier in section \ref{sec:massrenormalization}, atoms propagate with reduced mass $M_i-K_{i,i}$, so that the effective mass, which includes that of the wave function cloud, is the physical mass $M_i$ of the atom. This mass renormalization counteracts the artificial underestimation of the vibrational frequencies within Car-Parrinello \textit{ab-initio} molecular dynamics simulations.\cite{grossman2004towards, ong2010vibrational}

%=========================================================
\subsubsection{Dynamics for energy eigenstates}
%=========================================================
The wave function dynamics described so far requires that the matrix of Lagrange multipliers $\Lambda_{m,n}$ for the orthonormality constraint of the Kohn-Sham wave functions is Hermitian. Hermitian Lagrange multipliers will lead to an energy-conserving dynamics. The resulting wave functions are, however, not eigenstates of the Hamiltonian but, as discussed earlier, some more-or-less arbitrary superposition of them.

For dynamical occupations, the wave functions need to be eigenstates of the Kohn-Sham Hamiltonian. 
This can be achieved by a diagonalization in each time step, as suggested by Marzari.\cite{marzari97_prl79_1337} 

In the CP-PAW code, we proceed differently: the dynamics is modified such that the wave functions approach the eigenstates of the Hamiltonian. This is achieved by restricting the Lagrange multipliers for the orthonormalization to the lower triangle by setting $\Lambda_{m,n}=0$ for $n>m$. As a consequence, the first wave function approaches the lowest eigenstate. The second wave function minimizes its energy, while maintaining orthogonality to the first state and thus approaches the second-lowest eigenstate. Each wave function experiences the constraint forces from only the preceding wave functions, which results in a dynamics that becomes stable only when all states are eigenstates of the effective Hamiltonian. This dynamics is in principle no longer energy-conserving because the electronic forces driving the gauge transformation are not reflected in the potential energy.
However, over time, the Lagrange multipliers become nearly diagonal and therefore Hermitian, which recovers energy conservation, at least approximately. 

It is advisable to initiate the dynamics described above already with eigenstates of the current Kohn-Sham Hamiltonian.
This requires a one-time diagonalization in the basis of the current wave functions.

%=========================================================
\subsection{K-points}
%=========================================================
For crystalline solids, we exploit Bloch's theorem that
identifies the eigenstates of the effective Hamiltonian for a crystal with the eigenstates of the lattice translation operator.\cite{bloch1929quantenmechanik} 
These are known as Bloch states, which are a product of a wave function that has the translational symmetry of the crystal potential, and a phase factor $\exp(i\vec{k}\vec{r})$, where $\vec{k}$ is a k-point in the reciprocal unit cell. The wave functions are constructed for each k-point individually, but expectation values of an operator $\hat{A}$ must be recovered by a Brillouin-zone integration via 
\begin{eqnarray}
    \langle A \rangle&=&\frac{1}{V_G}\int_{V_G} d^3k\; 
   \nonumber\\&\times& \hspace{-3mm}
    \sum_n
    f_{T,\mu}\big(\epsilon_n(\vec{k})\big)
    \big\langle\psi_n(\vec{k})\big|
    \hat{A}\big|\psi_n(\vec{k})\big\rangle.
\end{eqnarray}
In order to evaluate this integral in practice, the Kohn-Sham energies $\epsilon_n(\vec{k})$ and the matrix elements $A_n(\vec{k})=\langle\psi_n(\vec{k})|\hat{A}|\psi_n(\vec{k})\rangle$ must first be evaluated for a discrete set of k-points and then interpolated to perform the integration.

One particular simple approach is the special-point scheme of Monkhorst and Pack:\cite{monkhorst76_prb13_5188}
if the integrand can be represented effectively by a Fourier interpolation, i.e. 
\begin{eqnarray}
f_{T,\mu}\big(\epsilon_n(\vec{k})\big)
A_n(\vec{k})=\sum_{\vec{t}}\exp(i\vec{k}\vec{t}) C^{(A)}_{n}(\vec{t}), 
\end{eqnarray}
with general lattice vectors $\vec{t}$, then the Brillouin-zone integration can be represented by the first term, hence
\begin{eqnarray}
\langle A \rangle = \sum_n
\underbrace{
\frac{1}{\sum_{\vec{k}}1} \sum_{\vec{k}} 
f_{T,\mu}\big(\epsilon_n(\vec{k})\big)
A_n(\vec{k})}_{C_n^{(A)}(0)}. 
\end{eqnarray}
For a sufficiently smooth integrand, this technique exhibits exponential convergence with the grid spacing, which is satisfactory.\cite{monkhorst76_prb13_5188} Important is, however, that the $k$-point set forms a regular grid, i.e. it forms itself a lattice in reciprocal space. 

For insulators at zero temperature, the occupations of the Kohn-Sham states are either one or zero for a complete band and are known beforehand.
The average over the regularly spaced k-point grid, called sampling, is the method of choice.

Yet, the situation is more complicated for metals because of the step-like behavior of the Fermi distribution function. Two different strategies are implemented in the CP-PAW code,
sampling with finite temperature occupations and the tetrahedron method,\cite{mermin65_pr137_A1441, alavi1994ab, richters2014self, jepsen71_ssc9_1763, lehmann72_pssb54_469, bloechl94_prb49_16223} both of which are described in the following.

%====================================================
\subsubsection{Mermin functional}
%===================================================
For metals at sufficiently high temperatures, the sampling technique is in principle also appropriate. However, a very fine k-point grid is required to resolve the step-like Fermi surface.
A common feature of metals is that their occupations need to be recalculated during the optimization or dynamics. Hence, 
\begin{eqnarray}
\Delta\mathcal{L}&=&\frac{1}{2}\sum_n
m_f \dot{x}_n^2
-k_BT\sum_n\Bigl[
f_n\ln\big(f_n\big)
\nonumber\\
&+&
(1-f_n)\ln\big(1-f_n\big)
\Bigr]
\end{eqnarray}
is added to the Lagrangian,\cite{gillan1989calculation} where the occupations $f_n=\frac{1}{2}\big(1-\cos(\pi x_n)\big)$ are a function of the dynamical variables $x_n(t)$.
These occupations are also part of the density functional, and the index $n$ identifies a specific Kohn-Sham wave function that may consist of a band index and a k-point.

As a cautionary remark, let us mention that the Euler-Lagrange equation for variable occupations needs to be adjusted to arrive at numerically stable equations of motion.

%========================================================
\subsubsection{Tetrahedron method}
%========================================================
The tetrahedron method uses the k-point grid to interpolate
energies and matrix elements.\cite{jepsen71_ssc9_1763,lehmann72_pssb54_469,bloechl94_prb49_16223} For this purpose, the reciprocal space is divided into tetrahedra so that their corners lie on the chosen k-points. The values at the four corners define a linear function that interpolates between the four values. In this manner, one obtains a piecewise analytical function for the Kohn-Sham energies and the matrix elements. For these piecewise functions, the integration is performed analytically. The modern extension of the tetrahedron method used in CP-PAW takes also the curvature of the matrix elements into account,\cite{bloechl94_prb49_16223} which dramatically improves the convergence with the number of k-points.

The integration is turned into an expression that appears like sampling, i.e.
\begin{eqnarray}
\langle A \rangle = \sum_n w_n(\vec{k}) A_n(\vec{k})
\end{eqnarray}
with integration weights $w_n(\vec{k})$ for the discrete k-point set, but is instead an interpolation followed by an analytical integration of the interpolated function.  The integration weights are determined for a set of energy values $\epsilon_n(\vec{k})$ and a specific particle number. 

Users often raise concerns about integration weights that lie outside the interval between zero and one. 
Yet, this is because the method interpolates the bands between the discrete k-points and then occupies these interpolated bands up to the Fermi surface with exactly one, respectively, two electrons. 
Hence, the interpolation weights must not be identified with occupations.

%\petertt{Retarded energy levels, Energy eigenstates as wave functions}

%====================================================
\subsubsection{Electrostatic decoupling}
%===================================================
\label{sec:decoupling}
Although CP-PAW has been developed with solids in mind, isolated molecules can be studied as well. 
The complication of plane-wave-based methods for molecules is the presence of periodic images:
Rather than a single molecule, a grid of molecules is simulated.
The remedy for this problem is to remove any interaction between these periodic images.

The overlap of wave function tails from different periodic images falls off exponentially and can be controlled by using larger lattice vectors.
Typically, a distance between periodic images in the range 6-10~{\AA} is sufficient. 
Even a small k-point grid, as opposed to just using a single k-point, may help to suppress bond formation between periodic images. 
However, because the electrostatic interaction is long-ranged, its interaction is more difficult to remove.
Several methods are used to achieve this.\cite{hockney1988computer, barnett1993born, bloechl95_jcp103_7422, makov95_prb51_4014, martyna1999reciprocal, genovese2006efficient}

In the CP-PAW code, we construct an auxiliary point charge model to accomplish the electrostatic decoupling of the periodic images.\cite{bloechl95_jcp103_7422}
The full charge distribution is converted into a point charge model in such a way that it reproduces the electrostatic potential far away from the molecule.
An Ewald summation for the periodic point charge model provides the energy in a lattice,\cite{ewald21_annp64_253} from which the energy of the isolated point charge model is removed.
Thus, the Coulomb interaction between periodic images is retrieved.
This total energy correction is incorporated into the Lagrangian, so that the back-action on the atom positions and wave functions is consistently taken into account.

For the sake of completeness, let us mention the role of a compensating charge background, which is treated consistently.
This term is particularly important when different charge states are compared such as for electron affinities or ionization potentials. 
We refer to the original publication for details.\cite{bloechl95_jcp103_7422}

%=================================================
\subsection{Local hybrid density functional} 
%=================================================
%\petertt{Can we keep this under a different heading than the PAW method?}
%\petertt{We need to explain the double counting!}
%\petertt{The motivation for using the local terms was not simplification, but screening.}
%\thomastt{Please double check!}
\label{sec:local_hybrid_functional}
Beside the popular generalized gradient approximation to the exchange-correlation energy $E_{\rm XC}$,\cite{becke1986density, perdew1986accurate} so-called hybrid density functionals,\cite{becke93_jcp98_1372} which include some amount of exact Hartree-Fock exchange, are also available in CP-PAW. 

Usual implementations of hybrid functionals in plane wave-based codes are computationally rather involved.\cite{todorova2006molecular, spencer2008efficient, carnimeo2019fast, mandal2019speeding} The local PBE0r hybrid functional employed in the CP-PAW code follows a different strategy:
the Kohn-Sham wave functions are first decomposed into localized atom-centered tight-binding orbitals $|\chi_{\mu}\rangle$ in which the exchange terms are then expressed. 

In the local PBE0r hybrid functional only the onsite four-center integrals are taken into account.
This accounts for the screening of the Coulomb interaction in the spirit of the GW approximation,\cite{hedin65_pr139_796} respectively, the random-phase approximation.\cite{bohm51_pr82_625}
The screening picks up the idea of range separation, as implemented, for example, in the HSE functionals.\cite{heyd03_jcp118_8207}

The limitation to onsite terms cures the well-known problems of density functionals with transition metal oxides, when their band gaps are grossly underestimated and are even erroneously predicted as metals.\cite{cohen2008insights} 
However, mitigating the band gap problem in covalent materials requires, in addition, the nearest-neighbor exchange terms, in the spirit of neglect of diatomic differential overlap techniques.\cite{gill00_jcomputchem21_1505}

Within this local approximation,\cite{bloechl11_prb84_205101, bloechl13_prb88_205139} the PBE0r functional contains local contributions of exact exchange
\begin{equation}
E_{\rm X,R}^{\rm HF} = -\frac{1}{2} \sum_{\mu, \nu, \lambda, \sigma \in C_R} \langle \mu \nu | \lambda \sigma \rangle \rho_{\lambda \nu}^{(1)} \rho_{\sigma \mu}^{(1)}, 
\end{equation}
where 
\begin{equation}
    \langle \mu \nu | \lambda \sigma \rangle = \iint d^3r d^3r' \, \frac{e^2\chi_\mu^* (\vec{r}) \chi_\nu^*(\vec{r'}) \chi_\lambda (\vec{r}) \chi_\sigma(\vec{r'})}{4\pi\epsilon_0 |\vec{r}-\vec{r'}|}
\end{equation}
are 4-center, 2-electron integrals, while $\rho_{\lambda \nu}^{(1)}$ and $\rho_{\sigma \mu}^{(1)}$ are one-particle reduced density matrices, as defined in Eq.~\ref{1-RDM}. $C_R$ is the set of orbitals centered at site $R$. The exchange terms contained already in the regular density functional $E^{PBE}_{\rm XC}$ are subtracted out in the double-counting term $E_{\rm X,R}^{\rm PBE}$. Contrary to the popular PBE0 hybrid functional,\cite{adamo1999toward} which includes 25\% of exact Hartree-Fock exchange, the local PBE0r hybrid functional is given by 
\begin{equation}
  E_{\rm XC}^{\rm PBE0r} = E_{\rm XC}^{\rm PBE} + \sum_R a_R \left( E_{\rm X,R}^{\rm HF} - E_{\rm X,R}^{\rm PBE}\right), 
  \label{eq:PBE0r}
\end{equation}
where $a_R$ is the element-specific fraction of exact Hartree-Fock exchange. 
Empirically determined values for $a_R$  are in the range of 10\%,\cite{roemer20_molecules25_5176} substantially smaller than the usual 25\% derived from the adiabatic connection formula.\cite{perdew96_jcp105_9982}

The core-valence exchange terms are explicitly included. % that is described elsewhere.\cite{bloechl11_prb84_205101}
%
%Since the latter is restricted to the onsite terms $E_{\rm X}^{\rm HFr}$, only the corresponding PBE exchange terms $E_{\rm X}^{\rm PBE0r}$ are subtracted in order to avoid double counting.
%
%The underlying idea of the double counting term of the PBE0r functional is to separate the Coulomb interaction into contributions of individual atoms using cutoff functions
%\begin{equation}
%  g_R(\vec{r}) = \frac{\rho_R(\vec{r})}{\sum \rho_{R'}(\vec{r})}, 
%\end{equation}
%where 
%\begin{equation}
%  \rho_R(\vec{r}) = \sum_{\mu \nu \in C_R} \sum_{\sigma} \langle \vec{r}, \sigma | \chi_{\mu} \rangle \rho^{(1)}_{\mu,\nu} \langle \chi_{\nu} | \vec{r}, \sigma \rangle
%\end{equation}
%are local electron-density contributions.\cite{bloechl11_prb84_205101} In practice, however, the double counting correction is approximated by evaluating both cutoff functions at the same position $\vec{r}$, which yields the simplified expression
%\begin{equation}
% E^{\rm PBE0r}_{\rm DC} = - \sum_R \int d\vec{r} \, \frac{\rho_R(\vec{r})}{\rho(\vec{r})} \epsilon_{\rm XC}(\vec{r}) \rho_R(\vec{r}).
%\end{equation}
%In the CP-PAW code, the electron density $\rho(\vec{r})$ is represented by means of a partial wave expansion, while $\rho_R(\vec{r})$ is expressed in terms of local orbitals. 

%===========================================
\section{Prerequisites}
%===========================================

%\petertt{Here we should include a docker file (container) so that readers can go through the installation without being concerned by the installation. We need to decide what to do with the description of the installation below. Remove, put into an appendix, combine with the manual?}
Requirements for the CP-PAW package are:
\begin{itemize}
\item Unix type operating system, e.g. Linux, MacOS
\item bash shell
\item Fortran compiler (Version 2008 compatible), e.g. gfortran.
\item Necessary libraries: LAPACK, BLAS, FFTW3, LibXC (optional), MPI (optional). Some of the libraries are part of packages such as ACML, MKL or Apple's framework "Accelerate". 
\item LaTeX with latexmk for the documentation
\item GNU make (version $\ge$4.3)
\end{itemize}

Some of the tools make use of external viewers such as xmgrace and gnuplot.

%=====================================================
\subsection{Installation}
%=====================================================
The CP-PAW package can be obtained at \url{https://github.com/cp-paw/cp-paw} from GitHub.
After downloading and unpacking, you will find an \verb|paw_install.sh| script in the top directory of the distribution. Executing it in this top directory, the installation script does its best to analyze your system and find the libraries. If successful, it will compile the manual and construct all executables.

The installation can be customized by adjusting the file
\url{src/Buildtools/defaultparmfile}. A revised copy of it
is used for the installation by passing it through the option -f to the \verb|paw_build.sh| script that is executed within the installation script.

For convenience, the resulting directories \verb|bin/dbg|, \verb|bin/fast| 
and \verb|bin/fast_parallel|
are added with the full path to the \verb|$PATH| variable inside the shell profile.

%=========================================
\subsection{The basic files of a CP-PAW project}
\label{sec:basicfiles}
%=========================================
There are four basic files for a simulation: 
\begin{itemize}
\item the \textit{"structure file"} (extension .strc),
\item the \textit{"control file"} (extension .cntl),
\item the \textit{"protocol file"} (extension .prot), and
\item the \textit{"restart file"} (extension .rstrt).
\end{itemize}
Other file types exist for special purposes, but will not be discussed here.

The \textit{"structure file"} describes what the system under study is, while the \textit{"control file"} specifies what one wants to do with it.
Structure and control files are input files supplied by the user.
The \textit{"protocol file"} provides information on the progress of the calculation and reports some basic physical data. At regular intervals, the code dumps the current state of the simulation into a \textit{"restart file"}, from which the calculation can be continued. 

The files of a CP-PAW project have a common rootname identifying the project and the extensions, which describe the content of the file.

%=========================================
\subsection{The input file format}
\label{sec:fileformat}
%=========================================
The input data, i.e. the \textit{"control and structure files"}, are organized in a fairly flexible format: The order of items and their positioning is irrelevant, as long as the logical order is preserved.

The data is represented in a logical tree structure. The basic elements are \textit{"branches"} and \textit{"leaves"}.
A \textit{"branch"} is a container, which may contain other \textit{"branches"} and \textit{"leaves"}. A \textit{"leaf"} is an actual data item. The logical identity of an object, \textit{"branch"} or \textit{"leaf"} is determined by its name and the sequence of branches to which it belongs. 

A \textit{"branch"} is enclosed by a starting tag, such as \verb|!CONTROL|, and the end tag \verb|!END|. The starting tag is the branch name -- in this case \verb|CONTROL| -- preceded by an exclamation mark.

A \textit{"leaf"} has only a starting tag, such as \verb|NSTEP=|, which is the name of the leaf -- in this case \verb|NSTEP| -- followed by an equal sign. The starting tag is followed by the data. The data section extends up to the next starting or closing tag of a \textit{"branch"}, or up to the starting tag of another \textit{"leaf"}.

The data types allowed are strings, logical, real values, and integer numbers. Data may be single or multiple values separated by spaces.
The data types are identified as follows:
\begin{center}
\begin{tabular}{ll}
\hline\hline
 identifier & data type\\
\hline
apostrophes (" or ') & string\\
T or F               & logical\\
period (.)           & real(float)\\
none of the above    & integer\\
\hline
\hline
\end{tabular}
\end{center}
All tags are case-insensitive.
The data structure ends with a closing tag \verb|!EOB|. All data following this closing tag are ignored. 

All keywords, their meaning, and their syntax are described in the manual \verb|doc/manual.pdf|. %The manual is accessed from the top directory as \verb|doc/manual.pdf|. 
The reader should make it a habit of consulting the manual for background information and to become familiar with the functionality of the CP-PAW code.

Unless otherwise specified, the data are given in Hartree atomic units $e=\hbar=m_e=4\pi\epsilon_0=1$ and Cartesian coordinates, which is the internal representation of the data in the code.

%=========================================
\subsection{The code structure}
%=========================================
%Most importantly, the syntax of all input data is described in the manual \verb|doc/manual.pdf|.

The CP-PAW package works with several layers:
The central part is the simulation engine. 
The analysis tools form the next layer. They use data written by the simulation engine in machine readable form and produce data files for inspection or as input for external viewers. 

The CP-PAW package also contains a number of shell scripts, which allow us to parse and manipulate data files of the package.
Particularly useful is the wrapper \verb|paw_do|, which packages shortcuts to a number of frequently used tasks. 
Inspect the options with \verb|paw_do -h|.

%=========================================
\section{Exercises}
%=========================================
For this tutorial, we have chosen the following as examples:
\begin{enumerate}
\item Malonaldehyde as an isolated molecule, 
\item Metallic iron to demonstrate solids and magnetism, and 
\item Praseodymium-manganite as an example of a strongly correlated insulator.
\end{enumerate}
Although the selected materials are simple, each has a few interesting properties that allow us to demonstrate the main functionality of CP-PAW. 
%\konstantintt{Is it an option to go to a one-column format to make the input files and commands more eligible to read?}

Input files and scripts for the automated execution of all exercises in this tutorial are available at \verb|https://github.com/cp-paw/tutorial|.
While we advise to install the code from source, we have a docker container available for easy installation to follow the exercises under \verb|https://github.com/cp-paw/container|.

%=======================================
\subsection{Exercise 1a: Wave function optimization}
\begin{figure}[!hbt]
    \centering
    \includegraphics[width=\linewidth,clip=true]{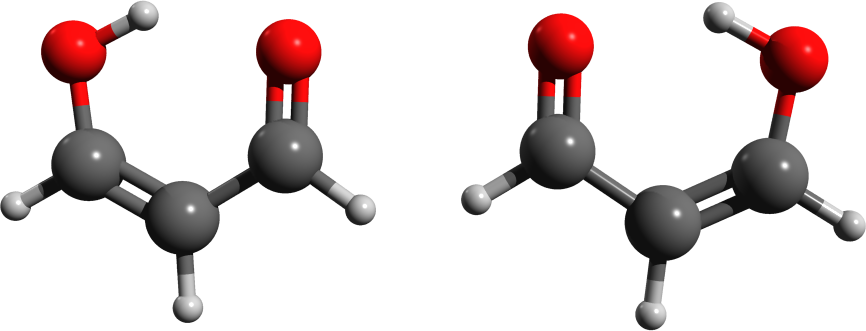}
    \caption{Two isomers of malonaldehyde before and after an intramolecular proton transfer.}
    \label{fig:twomalonaldehide}
\end{figure}

%=======================================
Malonaldehyde, in short O(CH)$_3$OH, exhibits an intramolecular proton transfer coupled to a soliton propagating along a short carbon chain.\cite{barone1996proton,wolf1998proton,wolf98_cej4_1418,list2020probing}
The two isomers of malonaldehyde are shown in Fig.~\ref{fig:twomalonaldehide}.

Before starting, the reader should be familiar with Sections~\ref{sec:basicfiles} and \ref{sec:fileformat}.
Two input files must be prepared, the \textit{"control file"} and the \textit{"structure file"}.

We will use the following \textit{"control file"} \verb|c3o2h4.cntl|:
%\verbatiminput{Inputfiles/c3o2h4.cntl_1}
\verbatiminput{Inputfiles/Peter2/src/Malo/c3o2h4_urlx_1.cntl}

\verb|START=T| tells the code to start from scratch, without looking for a \textit{"restart file"}. The value of \verb|NSTEP=| sets the number of iterations. The branch \verb|!PSIDYN!AUTO| contains suitable settings of the minimization scheme for the wave function optimization described in Section~\ref{sec:optimization}.

Let us now turn to the \textit{"structure file"}, \verb|c3o2h4.strc|:
{\footnotesize
\verbatiminput{Inputfiles/Peter2/src/Malo/c3o2h4.strc}}

The branch \verb|!OCCUPATIONS| specifies with \verb|EMPTY=5| that five empty states are calculated in addition to the filled states. 
It is advisable to include a few empty orbitals to see whether the molecule has developed a band gap. 
The number of empty orbitals can be adjusted later.

With \verb|LUNIT[AA]=1.|, a length unit of 1~{\AA} is specified for the atomic positions and lattice vectors, respectively.

Each atom is described by an \verb|!ATOM| block, specifying a name and the corresponding initial position in Cartesian coordinates. 
The name uniquely identifies the atom in the unit cell.
It begins with the two-letter element symbol, which, in turn, refers to the name of an atom type, described by a corresponding \verb|!SPECIES| branch. 
For elements with a one-letter symbol, an underscore is attached.

The lines with "\verb|@hydrogen@|", "\verb|@carbon@|" and "\verb|@oxygen@|" are place holders for the corresponding species files from appendix~\ref{sec:setupfiles}. The content of the species files needs to be inserted manually or using the shell script \verb|paw_resolve|.
The species branches describe the augmentation of the atom types and are provided in appendix~\ref{sec:setupfiles}.

Although we study an isolated molecule, we need to specify lattice vectors with \verb|T=| in the \verb|!LATTICE| block.
The lattice vectors are given as an array $\vec{T}_1,\vec{T}_2,\vec{T}_3$
with nine numbers in total. 
For molecules, a fcc-lattice is usually a good choice because it offers a good balance of high symmetry, a large distance between periodic images of the molecule, and a small unit cell volume.
The CP-PAW code actually simulates a grid of repeated molecules, which requires a distance of at least 6~{\AA} between the periodic images to ensure that the wave function has decayed to zero. 
The \verb|!ISOLATE| block subtracts the long-ranged Coulomb interaction between repeated molecules, as described in Section~\ref{sec:decoupling}.

The \verb|!CONSTRAINTS| branch suppresses a global translation and rotation of the molecule. 
This will be important later for the Car-Parrinello \textit{ab-initio} molecular dynamics simulation to prevent the aforementioned flying ice-cube effect.

%\paragraph{Wave function optimization:}
Now, we are ready to conduct our first optimization of the wave functions:  
\begin{verbatim}
paw_fast.x c3o2h4.cntl 1>out 2>&1 &
\end{verbatim}
For MPI installed, one can use instead
\begin{verbatim}
mpirun -np 10 --oversubscribe\
  $(which ppaw_fast.x) c3o2h4.cntl \
  1>out 2>&1 &
\end{verbatim}
The parameter \verb|-np| specifies the number of parallel processes executed and should reflect the number of cores on your computer.

The protocol file can be streamed to the command line with
  \begin{verbatim}
tail -f c3o2h4.prot
\end{verbatim}

The standard output and standard error are redirected to the file \verb|out|. 
This file is rarely relevant for the user.
At variance, the protocol file \verb|c3o2h4.prot| is important and monitors the execution.  

The user shall verify the convergence by inspecting the
"\textit{protocol file}", or the graph produced by executing \verb|paw_show -ce c3o2h4| within the project directory.

%=======================================
\subsection{Exercise 1b: Structure optimization}
%=======================================
In order to continue from the restart file written by the previous calculation,
the parameter \verb|START| in the control file is removed or set to false, i.e. \verb|START=F|.
The number of time steps is increased to, e.g. \verb|NSTEP=1000|. While the Car-Parrinello method may require many optimization steps, the computational effort for each step is relatively small.

The atoms are allowed to move by including a block \verb|!RDYN| in the control file. 
The optimization scheme for the wave functions, specified by the \verb|!AUTO| branch in the \verb|!PSIDYN| block, is adjusted and the one for the atoms is added:
\begin{verbatim}
!PSIDYN
  !AUTO FRIC(-)=0.01 FACT(-)=0.9
        FRIC(+)=0.20 FACT(+)=1.0
        MINFRIC=0.01 !END
!END
!RDYN
  !AUTO FRIC(-)=0.0 FACT(-)=0.97
        FRIC(+)=0.1 FACT(+)=1.0 !END
!END 
\end{verbatim}
Because atoms have much lower frequencies than the wave functions, their optimum friction according to Eq.~\ref{eq:aopt} is also lower.

The lower friction \verb|FRIC(-)=0.0| for the atoms is set to zero, because the atoms experience the wave-function friction indirectly through the strong coupling of atoms and wave functions. 
The rather large value \verb|FRIC(+)=0.1| of the upper atom friction effectively quenches the atomic motion each time the energy passes through a minimum. 

The structure optimization can be accelerated by preconditioning.
Hereby, the atom masses are adjusted to narrow down the frequency spectrum.
One may make hydrogen 2-3 times more heavy and one may reduce the masses of the other atoms to a smaller value such as 5-10 atomic mass units.
This does not affect the ground-state configuration, but it does affect the time scales of the dynamics.
Therefore, one must return to the physical masses before performing a Car-Parrinello \textit{ab-initio} molecular dynamics simulation. The code is started with the same command as before.

%=======================================
\subsection{Exercise 1c: Collect information}
%=======================================
%=======================================
%\subsubsection{Protocol file}
%=======================================
The \textit{"protocol file"} \verb|c3o2h4.prot|:
\begin{itemize}
\item records the setting of CP-PAW,
\item logs the convergence process, and
\item reports energies, atomic structure, and energy levels.
\end{itemize}
A record of unrecognized input items in the control and structure files is provided in the protocol file to identify mistyped keywords.

The \textit{"protocol file"} logs the convergence process step-by-step. 
The parameters listed are described in Table~\ref{tab:logging}.
\begin{table}[!hbt]
\begin{tabular}{ll}
\hline
NFI & iteration counter\\
T[PSEC] & time in picoseconds\\
T[K]  & atomic kinetic energy in Kelvin\\
EKIN(PSI) & wave function kinetic energy\\
E(RHO) & DFT total energy \\
ECONS & potential and kinetic energy\\
ANNEE & wave function friction parameter\\
ANNER & atom friction parameter\\
\hline
\end{tabular}
\caption{Sequential data written to the protocol file.}
\label{tab:logging}
\end{table}
 This logging information can also be inspected with the tool \verb|paw_show|. Explore its calling sequence with \verb|paw_show -h|.
The command \verb|paw_do -A| is a shorthand of \verb|paw_show -ce| \textit{rootname}. 

Typically, every hundred time steps, more detailed information is written about the current atomic structure, forces, energy contributions, and energy eigenvalues.
This information is produced also when the simulation terminates.

%=======================================
%\subsubsection{Structure tool}
\paragraph{Structure tool:}
%=======================================
The structure tool \verb|paw_strc| writes bond distances and bond angles into the file \textit{rootname}\verb|.sprot|. 
Furthermore, the atomic structure is written in several common file formats, such as xyz, cml\footnote{chemical markup language}, cssr\footnote{SERC Daresbury Laboratory's Cambridge Structure Search and Retrieval file format.} to be used with the molecular viewer of your choice. 

The command \verb|paw_do -S| is a shorthand for \verb|paw_strc -cd|~\textit{rootname}, the typical call of the structure tool.

%=======================================
\paragraph{Density of States}
%=======================================
The projected density of states method is a powerful tool to analyze chemical bonding. We define the density of states operator as
\begin{eqnarray}
  \hat{D}(\epsilon)=\sum_n\int \frac{d^3k}{(2\pi)^3}
  |\psi_n(\vec{k})\rangle \delta(\epsilon-\bar{\epsilon}_n(\vec{k}))\langle\psi_n(\vec{k})|.
\end{eqnarray}

This density of states operator can be represented in terms of local orbitals using the decomposition
\begin{eqnarray}
|\psi_n(\vec{k})\rangle=\sum_{\alpha}|\chi_\alpha\rangle\langle\pi_\alpha|\psi_n(\vec{k})\rangle,
\end{eqnarray}
where $|\psi_n(\vec{k})\rangle$ are the Kohn-Sham wave functions and $|\chi_\alpha\rangle$ the local orbitals. Moreover, $|\pi_\alpha\rangle$ are the projector functions for the local basis, which satisfy the bi-orthogonality condition $\langle\pi_\alpha|\chi_\beta\rangle=\delta_{\alpha,\beta}$.
The projector function probes the character of a specific local orbital in a wave function.
The decomposition eventually yields 
\begin{eqnarray}
    \hat{D}(\epsilon)=\sum_{\alpha,\beta}|\chi_\alpha\rangle 
    D_{\alpha,\beta}(\epsilon)\langle\chi_\beta|.
\end{eqnarray}

From this operator, we may derive the total density of states
\begin{eqnarray}
    D_{tot}(\epsilon)=\textrm{Tr}[\hat{D}(\epsilon)],
\end{eqnarray}
a projected density of states that is projected onto an orbital $|\chi_\alpha\rangle$, i.e.
\begin{eqnarray}
    D_{\alpha,\alpha}(\epsilon)=\langle\pi_\alpha|\hat{D}(\epsilon)|\pi_\alpha\rangle,
\end{eqnarray}
or an off-diagonal density of states matrix element
\begin{eqnarray}
    D_{\alpha,\beta}(\epsilon)=\langle\pi_\alpha|\hat{D}(\epsilon)|\pi_\beta\rangle,
\end{eqnarray}
which may be called crystal orbital populations.

These are related to the well-known crystal orbital overlap populations\cite{hughbanks1983chains}
\begin{eqnarray}
COOP_{\alpha,\beta}(\epsilon)
=D_{\alpha,\beta}(\epsilon)\langle\chi_\beta|\chi_\alpha\rangle
\end{eqnarray}
after multiplication with an overlap matrix element, as well as with the crystal orbital Hamilton populations\cite{dronskowski93_jpc97_8617,lucke2017efficient}
\begin{eqnarray}
COHP_{\alpha,\beta}(\epsilon)=D_{\alpha,\beta}(\epsilon)
\langle\chi_\beta|\hat{h}_{eff}|\chi_\alpha\rangle
\end{eqnarray}
after multiplication with a Hamilton matrix element. 

The expectation values of one-particle operators can be obtained from the Fermi distribution function $f_{T,\mu}(\epsilon)=1/(\e{(\epsilon-\mu)/(k_BT)}+1)$, the density of states, and the matrix elements $A_{\alpha,\beta}=\langle\chi_\alpha|\hat{A}|\chi_\beta\rangle$ of the operator $\hat{A}$ as
\begin{eqnarray}
 \langle A\rangle=\sum_{\alpha,\beta}   
 \int d\epsilon\; f_{T,\mu}(\epsilon) D_{\alpha,\beta}(\epsilon)A_{\beta,\alpha}.
\end{eqnarray}
%The off-diagonal matrix elements of the density of states turn into the known crystal orbital overlap populations (COOP) after multiplication with the corresponding overlap matrix element, or into the crystal orbital Hamilton populations (COHP) after multiplication with a matrix element of the Hamiltonian.\cite{dronskowski93_jpc97_8617,lucke2017efficient}

For the analysis of chemical binding, it is very useful to be able to construct arbitrary composite orbitals and to inspect their projected density of states or crystal orbital populations, i.e.
\begin{eqnarray}
\langle\Phi|\hat{D}|\Psi\rangle
=\sum_{\alpha,\beta}
\langle\Phi|\chi_\alpha\rangle
D_{\alpha,\beta}(\epsilon)
\langle\chi_\beta|\Psi\rangle.
\end{eqnarray}
The CP-PAW code provides a means of specifying arbitrary superpositions of the primitives as composite orbitals.

Currently, the primitives are constructed from the partial waves used for the PAW augmentation.
These partial waves are truncated at some user-defined radius and orthonormalized on each site using a Cholesky decomposition of the local overlap matrix. 

The density of states is constructed in two stages:
\begin{enumerate}
    \item The relevant density of states matrix elements are chosen using the \verb|paw_dos| tool.
    \item  The \verb|paw_dosplot| tool selects specific density of states and arranges them in a plot. 
\end{enumerate}
The essentials of the \textit{"dos control file"} \verb|c3o2h4.dcntl| for the \verb|paw_dos| tool look as follows:

{\footnotesize
\begin{verbatim}
!DCNTL
  !GENERIC PREFIX=’Dos/’ !END
  !GRID BROADENING[K]=2000. !END
  !WEIGHT ID=’total’ TYPE=’TOTAL’ !END
  !WEIGHT ID=’hs’
    !ATOM NAME=’H_4’ TYPE=’ALL’ !END
    !ATOM NAME=’H_7’ TYPE=’ALL’ !END
    !ATOM NAME=’H_8’ TYPE=’ALL’ !END
    !ATOM NAME=’H_9’ TYPE=’ALL’ !END
  !END
!END
!EOB
\end{verbatim}
}

%\petertt{Shall we instead select the oxygen Dos and the OH coop? Thew}

The prefix places the files produced by the \verb|paw_dos| tool into the directory \verb|Dos|, which needs to be created beforehand. 

Because the density of states of molecules is a sum of delta functions, a thermal smearing is applied to produce a continuous density of states. 
The smearing is specified by the parameter \verb|BROADENING[K]=2000.|, which corresponds to a temperature of 2000~K. 
The default is room temperature, which is suitable for solids.
The density of states is constructed using the blocks \verb|!WEIGHT| for the diagonal and \verb|!COOP| for the off-diagonal matrix elements of the density of states, respectively. 
The \verb|!WEIGHT| block produces a file specified by the \verb|ID|, such as \verb|Dos/hs.dos| for \verb|ID=hs| and prefix \verb|Dos/|.
It sums up the contributions of all the \verb|!ATOM| blocks. % contained. 

Each density of states function is composed of pre-defined selections. 
One may select individual orbitals, or subsets of angular momentum weights, for a specified atom. 
For the block \verb|!ATOM|, one may select \verb|TYPE=| as 'S', 'P', 'D', 'F' or 'ALL', respectively.

Composite orbitals are defined via \verb|!ORBITAL| from primitives or other composite orbitals.
The primitives are tight-binding orbitals of cubic harmonics, such as 'S', 'PX', 'PY', 
'PZ', 'D3Z2-R2', etc., or the common hybrid orbitals
'SP1', 'SP2' and 'SP3', respectively. 

For each atom, a local coordinate system can be defined. 
The orientation of the axes can be defined either in Cartesian coordinates or using neighboring atoms.
The orbitals can be placed on arbitrary atomic positions in the first unit cell or any other unit cell.
The composite orbitals themselves can be superimposed to form even more complicated orbitals. 

Once the densities of states are defined, they need to be arranged using the \verb|paw_dosplot| tool, which in turn produces the input for the graphics tool \verb|xmgrace|.
 
The \verb|paw_dosplot| tool uses the \textit{"dosplot control file"} \verb|c3o2h4.dpcntl|, i.e.  
{\footnotesize
\begin{verbatim}
!DPCNTL
  EMIN[EV]=-35. EMAX[EV]=3. 
  YMIN=-0. YMAX=4.
  !GRAPH PREFIX=’Dos/’
    !SET ID=’total’ COLOR=’BLACK’ !END
    !SET ID=’h’ COLOR=’BLUE’ !END
    !SET ID=’c’ COLOR=’GREEN’ STACK=T !END
    !SET ID=’o’ COLOR=’RED’   STACK=T !END
  !END
  !GRAPH PREFIX=’Dos/’ 
    YMIN=0. YMAX=2. SCALE=1.0
    !SET ID=’os’ COLOR=’YELLOW’ !END
    !SET ID=’op’ COLOR=’RED’ STACK=T !END
  !END
!END
!EOB
\end{verbatim}
}
%\petertt{We need a figure here...}

%
The output is a set of graphs, each of which may contain several datasets, which are distinguished by their colors. 

The projected density of states can be represented in an intuitive manner by filling the graphs with a specific color and stacking them on top of each other. 
This gives an instant impression on the relative importance of the various contributions. 
By setting the parameter \verb|STACK=T|, the dataset can be stacked on top of the previous one, rather than counting the function value from the baseline.

Once the control files are prepared, they may be executed with the shorthand \verb|paw_do -D|, which executes:
\begin{verbatim}
paw_dos rootname.dcntl
paw_dosplot rootname.dpcntl
xmgrace -free -noask -batch rootname.bat 
\end{verbatim}

%=======================================
%\subsubsection{Wave functions}
%=======================================
%=======================================
%\subsubsection{paw\_get}
%=======================================

%=====================================================
\subsection{Exercise 2: \textit{Ab-initio} molecular dynamics}
%=====================================================
Let us now perform a Car-Parrinello \textit{ab-initio} molecular dynamics simulation of malonaldehyde at room temperature. We will observe intramolecular proton transfers between the two oxygen ions.
%and the correlated flipping of the double-bond network. The latter may be interpreted as a soliton moving back and forth the backbone.

The starting point is the \textit{"restart file"} obtained after the structure optimization of malonaldehyde. 
The time step, however, is set to a large number, e.g. \verb|NSTEP=20000|, though the calculation can be interrupted and continued as desired.

%\subsubsection{Thermostating the wave functions and atoms}
\paragraph{Molecular dynamics and thermostats:}
Rather than specifying an \verb|!AUTO| minimization scheme for the atomic and wave function dynamics, we specify thermostats. 
Hence, the \textit{"control file"} contains the following two branches: 
\begin{verbatim}
!PSIDYN 
  !THERMOSTAT STOP=T FREQ[THZ]=100. !END
!END
!RDYN RANDOM[K]=300.
  !THERMOSTAT STOP=T T[K]=300. 
     FREQ[THZ]=2.5 !END
!END
\end{verbatim}

As described earlier, the thermostat for the wave function dynamics is not a thermostat in the normal sense. 
Instead of imposing a temperature, it counteracts the unavoidable heat transfer from the "hot" nuclear degrees of freedom to the "cold" wave function degrees of freedom.\cite{bloechl92_prb45_9413,bloechl02_prb65_104303} 

For the atomic motion, we use a Nos\'e-Hoover thermostat, which produces a canonical ensemble. \cite{nose84_molphys52_255,hoover85_pra31_1695} 
With \verb|T[K]=300.|, the temperature is set to 300~K.

The Nos\'e-Hoover thermostat has a characteristic time scale for the energy fluctuations, which is set with \verb|FREQ[THZ]=2.5| to 2.5~THz.
Initially, this frequency is placed best in the center of the vibrational density of states, to achieve fast equilibration.
%Another tool to reach rapid thermalization is a friction acting on the thermostat variables.
During the data accumulation, however, the dominant frequency of the thermostat should lie below the frequencies of interest.

Before the system is equilibrated, the Nos\'e-Hoover thermostat exhibits large temperature fluctuations, which may even destroy the molecule of interest.
As a remedy, we provide a random kick to the atoms, which brings the system closer to thermal equilibrium. 
The parameter \verb|RANDOM[K]=300.| provides a kinetic energy of 300~K.

Initially, the energy is distributed over only a few vibrational modes, while with time, anharmonic effects distribute the energy evenly over all degrees of freedom.
Since small systems have larger relative energy fluctuations, the initial energy fluctuations exceed those in thermal equilibrium.

The system will need some time to equilibrate, until it represents a canonical ensemble, and before it can be used for data accumulation. Monitoring equipartition is a recommended test to verify equilibration.

Since the internal degrees of freedom are not coupled to the global translational and rotational motion, the latter are suppressed in the \textit{"structure file"} by including
\begin{verbatim}
!CONSTRAINTS
  !TRANSLATION !END
  !ROTATION !END
!END
\end{verbatim}

If the atomic masses had been adjusted to accelerate convergence, they must be restored now to the physical values to obtain the correct time scales. The isotope-averaged masses are the default values.

%\subsubsection{Analysing molecular dynamics trajectories}
\paragraph{Analyzing molecular dynamics trajectories:}
The trajectory can be analyzed with the \verb|paw_tra| tool, which uses its own \textit{"trajectory control file"}.
Similar to our density of states tool, the trajectory tool works in a modular fashion and allows for constructing complex vibration modes from simpler ones. 

For the purpose of demonstration, let us first inspect the internal hydrogen transfer. 
Specifically, we will observe the distances of the hydrogen atom \verb|'H_8'| from the two oxygen partners \verb|'O_5'| and \verb|'O_6'|, respectively, and we will demonstrate the arrangement of the double-bond network of the backbone as an example of a composite mode. 

To this end, we first define a mode named \verb|'O5-H8'|, which consists of the intramolecular bond distance between the atoms \verb|'O_5'| and \verb|'H_8'|. With the branch \verb|!OUTPUT|, this mode is assigned to the output file \verb|o5-h8.dat|. The corresponding time derivatives are plotted into \verb|o5-h8.vdat|. Similarly, the mode \verb|'O6-H8'| is also defined. 
%A minimal example for a \textit{"trajectory control file"}  \verb|c3o2h4.tcntl| is given below:
%{\scriptsize
%\verbatiminput{Inputfiles/Peter2/Snippets/c3o2h4.tcntl_minimal}
%}
The resulting \textit{"trajectory control file"} \verb|c3o2h4.tcntl| is given below:
{\scriptsize
\begin{verbatim}
!TCNTL
  !MODE ID='O5-H8'
    !BOND ATOM1='O_5' ATOM2='H_8' SCALE=1. !END
  !END
  !OUTPUT ID='O5-H8' FILENAME='o5-h8.dat' !END
  !OUTPUT ID='O5-H8' FILENAME='o5-h8.vdat' 
          TYPE='VELOCITY' !END
  
  !MODE ID='O6-H8'
    !BOND ATOM1='O_6' ATOM2='H_8' SCALE=1. !END
  !END
  !OUTPUT ID='O6-H8' FILENAME='o6-h8.dat' !END
!END
!EOB
\end{verbatim}
}

%First, we define the modes before connecting them to the corresponding output files.

%{\scriptsize
%\verbatiminput{Inputfiles/Peter2/Snippets/c3o2h4.tcntl_minimal}
%}

%The corresponding result is shown as black and red curves in Fig.~\ref{fig:malodistances}. 

\begin{figure}[!hbt]
\begin{center}
\includegraphics[width=\linewidth]{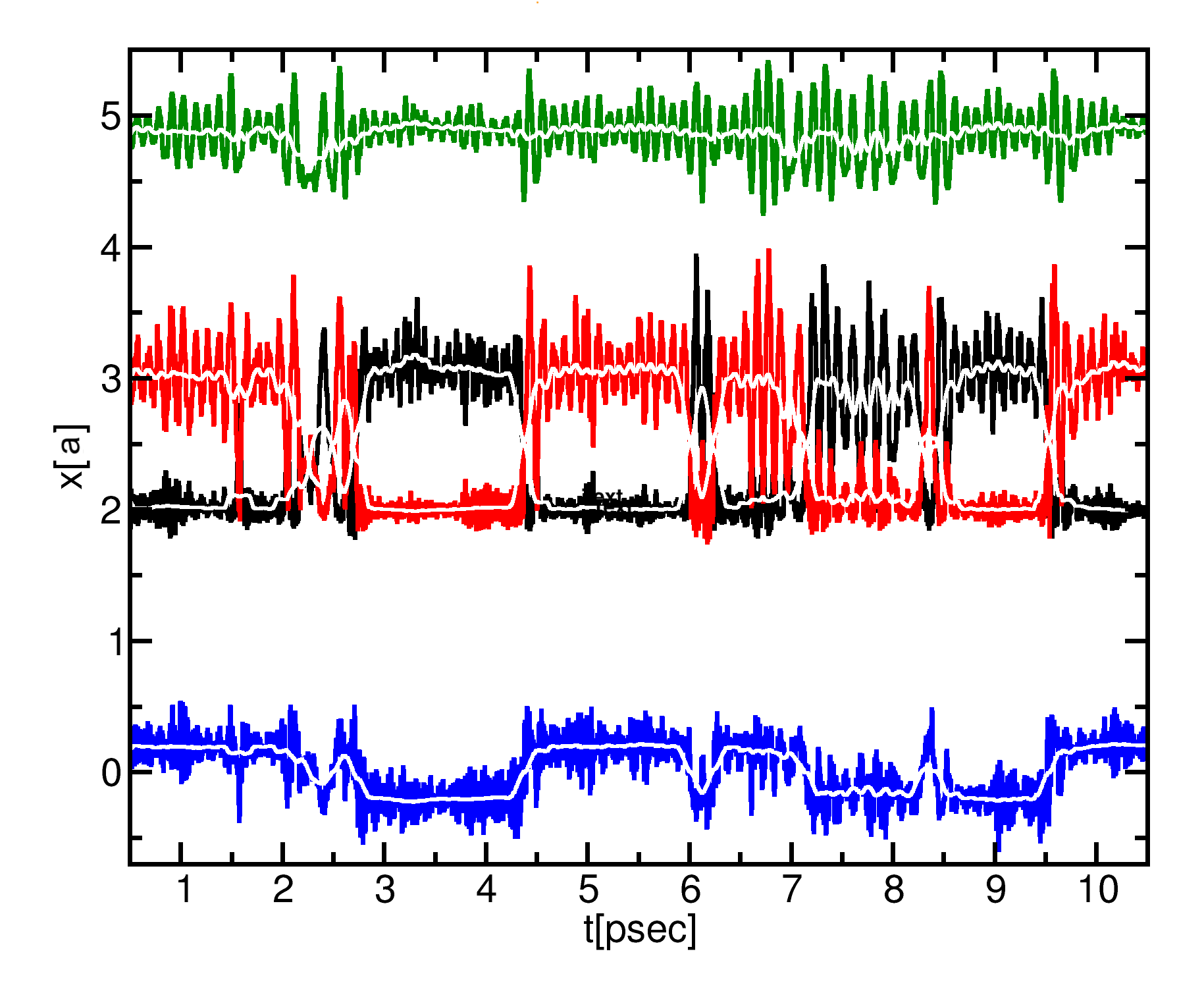}
%{data/fig5.png}
\end{center}
\caption{Long and short OH bond distances (red and black) in $a_{B}$ as a function of time in ps for malonaldehyde. For comparison, the double-bond network (blue) and intermolecular O-O distance (green) are also shown. %red and black: intramolecular OH distances, green: intermolecular O-O distance blue: double-bond network.
}
\label{fig:malodistances}
\end{figure}

It can be executed using the \verb|paw_tra| tool, and its output displayed via 
%Now we execute the \verb|paw_tra| tool and inspect its results
\begin{verbatim}
paw_tra c3o2h5.tcntl
xmgrace -free -noask o5-h8.dat o6-h8.dat
\end{verbatim}
The corresponding result, containing the two intramolecular OH distances of the hydrogen bond, is shown as black and red curves in Fig.~\ref{fig:malodistances}. Therewith, one can also see how the long and short intramolecular OH bonds interchange as the hydrogen atom hops between sites. 

We also find that the hydrogen atom is always covalently bonded to one of the two oxygen atoms and performs high-frequency bond distance oscillations with a period of 12.9~fs, which corresponds to 2586~$cm^{-1}$. 
In this case, one can simply monitor the velocity of the bond distance and read of the time delays between its zeros. %In Fig.~\ref{fig:malodistances}, one can also see the two intramolecular OH distances of the hydrogen bond and how the long and short intramolecular OH bonds interchange as hydrogen atoms hop between sites. 

%\thomastt{The sign of the mode de...}

To gain some more insights into this process, the intermolecular O-O mode, measuring the distance between the atoms \verb|'O_5'| and \verb|'O_6'|, is added to our \verb|tcntl|-file. 
In addition, the double-bond network mode \verb|'DBN'| is constructed from the four bond distances of the backbone with alternating sign, as specified by the \textit{"leaf"} \verb|SCALE=| within the \verb|tcntl|-file, i.e.
\begin{eqnarray}
    x_{dbn}=d_{O_5,C_1}-d_{O_1,C_2}+d_{C_2,C_3}-d_{C_3,O_6}.
    \label{eq:dbn}
\end{eqnarray}

Both is accomplished by inserting the following block into the file \verb|c3o2h4.tcntl|: 
{\scriptsize
\begin{verbatim}
  !MODE ID='O5-O6'
    !BOND ATOM1='O_5' ATOM2='O_6' SCALE=1. !END
  !END
  
  !MODE ID='DBN'
    !BOND ATOM1='O_5' ATOM2='C_1' SCALE=+1. !END
    !BOND ATOM1='C_1' ATOM2='C_2' SCALE=-1. !END
    !BOND ATOM1='C_2' ATOM2='C_3' SCALE=+1. !END
    !BOND ATOM1='C_3' ATOM2='O_6' SCALE=-1. !END
  !END
  !OUTPUT ID='DBN'   FILENAME='dbn.dat' !END
\end{verbatim}
}

%The result is the blue curve in Fig~\ref{fig:malodistances} ...

In Fig.~\ref{fig:malodistances}, the length of the hydrogen bond in terms of the intermolecular O-O distance is now shown in green. Hence, the intramolecular OH bond distance (red and black), and eventually also the frequency of the OH vibration, are strongly correlated with the length of the hydrogen bond.

Moreover, the double-bond network of the backbone, which is shown in blue, is correlated with the position of the hydrogen atom. 
This indicates a soliton moving along the backbone as the hydrogen switches its oxygen partners.
One can also observe how fluctuations with short hydrogen bond distances trigger attempted jumps of the hydrogen atom.

In order to ensure that the system is in thermal equilibrium, we may test the equipartition theorem.  
In order to obtain the temperature for various groups of atoms, we insert the following block into the \verb|tcntl|-file:
{\footnotesize
\begin{verbatim}
!TEMPERATURE RETARD[PS]=1.0
  !FILE EXT=T NAME='_t_all.dat' !END
!END
!TEMPERATURE RETARD[PS]=1.0
  !SELECT ATOMS='H_8' !END
  !FILE EXT=T NAME='_t_h8.dat' !END
!END
!TEMPERATURE RETARD[PS]=1.0
  !SELECT ATOMS='H_9' 'H_4' 'H_7' 'H_8' !END
  !FILE EXT=T NAME='_t_h.dat' !END
!END
!TEMPERATURE RETARD[PS]=1.0
  !SELECT ATOMS= 'C_1' 'C_2' 'C_3' !END
  !FILE EXT=T NAME='_t_c.dat' !END
!END
!TEMPERATURE RETARD[PS]=1.0
  !SELECT ATOMS='O_5' 'O_6' !END
  !FILE EXT=T NAME='_t_o.dat' !END
!END
\end{verbatim}
}
With \verb|retard[PS]| we specify a time scale $\tau$ for a running average by means of
\begin{eqnarray}
gk_BT(t)=\int_{-\infty}^t dt'\; 
{\rm e}^{-(t-t')/\tau} E_{kin}(t'),
\end{eqnarray}
where $g$ is the number of vibrational degrees of freedom in the selection, $k_B$ is Boltzmann's constant, and $E_{kin}$ is the kinetic energy of the atoms in the selection. 
However, caution is required, because the conversion into the temperature does not subtract the translational and rotational degrees of freedom. 
Hence, the data must be scaled up
by a factor ${3N}/{3N-6}\approx 1.28$, where $N=9$ is the number of atoms.

\begin{figure}[!hbt]
\begin{center}
\includegraphics[width=\linewidth]{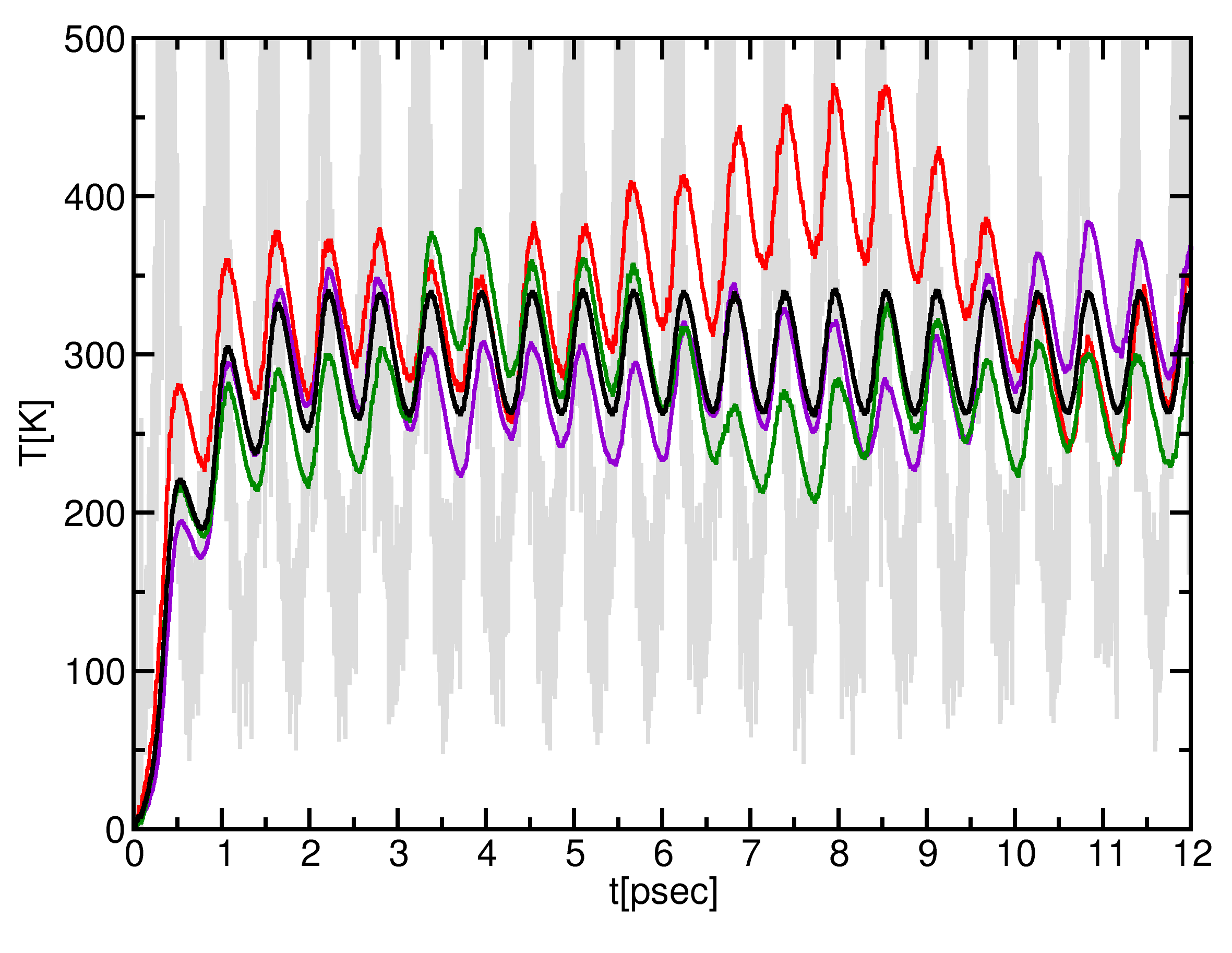}
%{data/fig6.png}
\end{center}
\caption{Temperature in Kelvin as a function of time in ps for all atoms. The true fluctuations (gray) are strongly suppressed by using a running average over 1~ps (black). Also shown are the averaged temperatures of the four hydrogen atoms (violet), the two oxygen atoms (red) and the three carbon atoms (green).}
%xmgrace c3o2h4_t_all.dat c3o2h4_t_h.dat c3o2h4_t_o.dat c3o2h4_t_c.dat
\label{fig:equipartition}
\end{figure}

The temperatures as a function of time are shown in Fig.~\ref{fig:equipartition}.
The running average over about 1~ps filters out the atomic oscillations, which occur on a sub-picosecond time scale. 
The visible oscillations of the total temperature are induced by the thermostat variable.
We can recognize some deviations from equipartition, with hot oxygen and cold hydrogen atoms.
These deviations serve as an error estimate of the temperature.

%=====================================================
\subsection{Exercise 3: Solids}
%=====================================================
For solids, some new aspects become relevant in comparison to the previous example of a molecular system.
In the current exercise, where we have chosen iron as an example, you shall become familiar with
\begin{itemize}
\item Brillouin-zone integration
\item variable occupations
\item magnetism
\item unit-cell optimization
\end{itemize}

%For this exercise we have chosen iron as an example.
Iron is a metal with a ferromagnetic ground state and a body-centered cubic (bcc) structure called ferrite ($\alpha$-iron). 
Ferrite undergoes a pressure-induced phase transition into the antiferromagnetic hexaferrum ($\epsilon$-iron) phase,\footnote{Hexaferrum is also a mineral, which contains amounts of other elements.\cite{mochalov98_zvmo127_41,cabri18_mineralogmag82_531}}
which will be the topic of this exercise. Here, we will study the non-magnetic phase of hexaferrum, as opposed to the physical antiferromagnetic phase.

Let us first set up the \textit{"structure file"} with the bcc structure of ferrite and a lattice constant of $a_{lat}=2.860$~\AA.\cite{seki05_isij45_1789}
% Ichiro Seki and Kazuhiro Nagata. Lattice constant of iron and austenite including its supersaturation phase of carbon. ISIJ Int., 45:1789, 2005.
%
The lattice vectors are $\vec{T}_1=(-\frac{1}{2},\frac{1}{2},\frac{1}{2})$~$a_{lat}$, $\vec{T}_2=(\frac{1}{2},-\frac{1}{2},\frac{1}{2})$~$a_{lat}$ and $\vec{T}_3=(\frac{1}{2},\frac{1}{2},-\frac{1}{2})$~$a_{lat}$, respectively.
For solids, we need to suppress the translation to avoid the previously described flying ice-cube effect. 
However, in contrast to molecules, we must not constrain the orientation of the atoms in the unit cell. Rotations of a solid are described by the lattice vectors, and the atomic positions must be able to adjust to the lattice vectors.

\paragraph{k-Points:}
When we study metals, we need to integrate over the k-points in the reciprocal unit cell.
This is called Brillouin-zone integration.
For this purpose, we choose a discrete grid of k-points. 
The grid spacing $\delta k\le\frac{2\pi}{R}$ is controlled by the parameter "R". 
Thus, we add the line 
\begin{verbatim}
!KPOINTS R=40. !END
\end{verbatim}
to the \textit{"structure file"} \verb|ferrite.strc|.

In a metal, the occupations need to be adjusted to the instantaneous band structure. 
Therefore, the occupations are specified as dynamical variables by including the branch \verb|!MERMIN| in the \textit{"control file"}:
\begin{verbatim}
!MERMIN START=T TETRA+=T T[K]=0. 
        ADIABATIC=T RETARD=20. !END  
\end{verbatim}
With \verb|TETRA+=T| we select the "improved" tetrahedron method,\cite{bloechl94_prb49_16223} which interpolates the bands between grid points and subsequently integrates the interpolated bands analytically up to the Fermi level.
With \verb|ADIABATIC=T|, we specify a retarded scheme to occupy the one-particle states:  the occupations are determined from a band structure that adiabatically follows the instantaneous band structure. 
The parameter \verb|RETARD=20.| specifies an exponential decay over 20 time steps.

\paragraph{Variable occupations:}
When the occupations are dynamical variables, they are chosen in accordance with the energy levels. 
A subtlety arises because the Car-Parrinello method does not diagonalize a Kohn-Sham Hamiltonian, but only optimizes the one-particle-reduced density matrix. 
Therefore, the energy levels are not immediately available, and the dynamics need to be tweaked so that they converge to energy eigenstates. 
This is done by setting the parameter \verb|SAFEORTHO=F| in the \verb|!PSIDYN| block of the control file to false.
We call this mode \textit{"unsafe"}, because it violates strict energy conservation when the wave functions deviate from energy eigenstates. 
However, strict energy conservation is only relevant for dynamical simulations.

When optimizing wave functions with \verb|SAFEORTHO=F|, it is  
important not to suppress the required electronic relaxation towards the energy eigenstates. 
Without a potential energy term for these driving forces, the energy remains constant or may even rise.

However, with \verb|SAFEORTHO=F|, the electronic dynamics can lead to a conflict with the wave function optimization scheme defined by \verb|!CONTROL!PSIDYN!AUTO|.
%The optimization scheme may attempt to suppress the dynamics, that is, however, required to reach convergence.
One cure is to disable the convergence scheme and use a fixed friction parameter.
Yet, computationally more efficient is to select a one-time initial diagonalization of the Hamiltonian using the parameter \verb|STRAIGHTEN=T|.
%\sanar{The entire SAFEORTHO discussion above appears a bit scattered. }\petertt{better?}

\paragraph{Magnetism:}
Magnetic systems are selected with the parameter \verb|NSPIN| in the structure input file, i.e.
\begin{verbatim}
!OCCUPATIONS
  EMPTY=8 NSPIN=2 SPIN[HBAR]=2. 
!END
\end{verbatim}
The parameter \verb|NSPIN=2| allows for magnetic systems with collinear spin densities, while the default \verb|NSPIN=1| corresponds to a non-spin-polarized calculation.
Non-collinear spin distributions are also possible with \verb|NSPIN=3|.
The initial spin is set to a finite value with \verb|SPIN[HBAR]=2.| in order to avoid global spin-reversal symmetry of the initial state.
In case of variable occupations, the occupations will be adjusted during the optimization.
With \verb|EMPTY=8| we include a sufficient number of empty states (with respect to an insulating non-spin-polarized state), so that all states below the Fermi level are represented by wave functions.

\paragraph{Electronic structure of ferrite:}
Now we are ready to optimize the electronic wave functions of ferrite.

\begin{figure}[!hbt]
  \centering
  \begin{subfigure}{\linewidth}
    \centering
    \includegraphics[width=\linewidth,clip=true]{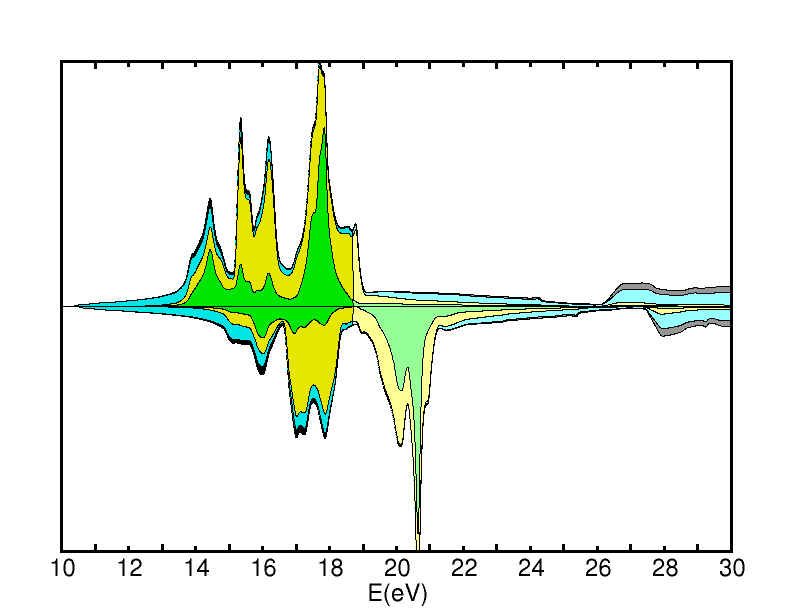}
    \caption{Ferrite}
    \label{fig:dosferrite}
  \end{subfigure}
  \vspace{0.5em}
  \begin{subfigure}{\linewidth}
    \centering
    \includegraphics[width=\linewidth,clip=true]{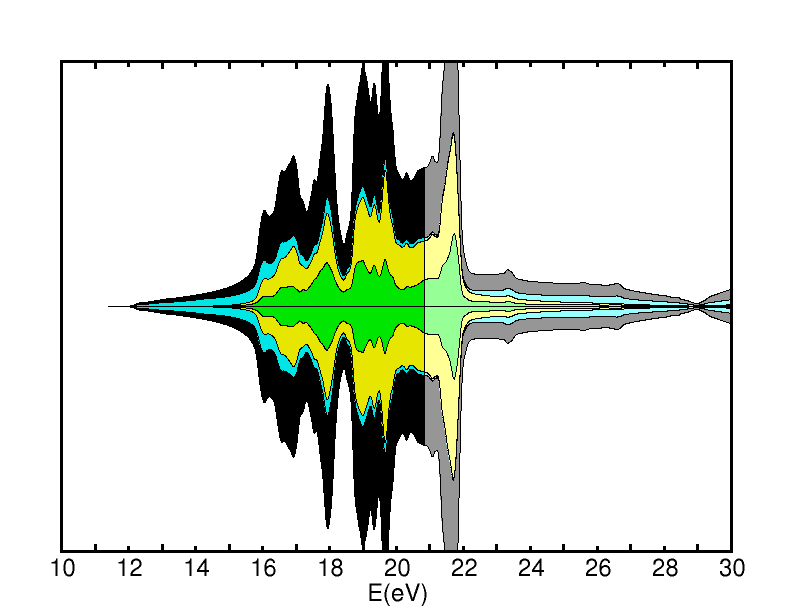}
    \caption{Hexaferrum}
    \label{fig:doshexaferrum}
  \end{subfigure}
  \caption{\label{fig:dos_solids}
  Spin-resolved density of states (a) ferrite and (b) hexaferrum, respectively. %with spin-down being plotted with negative sign.
  The spin-down density states is plotted with negative sign.
  Shown are the total density of states (black) and orbital projections:
  e$_g$ (green), t$_{2g}$ (yellow), and s/p (cyan) orbitals of iron.
  }
\end{figure}

Let us first inspect the density of states of ferrite, which is provided in Fig.~\ref{fig:dosferrite}. 

% \begin{figure}[!hbt]
% \begin{center}
% \includegraphics[width=\linewidth,clip=true]
% %{Inputfiles/Peter2/Figs/Fig7/ferrite_dos.png}
% {data/fig7a.png}
% \includegraphics[width=\linewidth,clip=true]
% %{Inputfiles/Peter2/Figs/Fig7/hexaferrum_dos.png}
% %{Figs/iron/ferrite_magnetic_dos.png}
% {data/fig7b.png}
% \end{center}
% \caption{Spin resolved density of states of ferrite (top) and hexaferrum (bottom). The spin down density states is plotted with negative sign. Total density of states (black) and a stack of e$_g$ (green), t$_{2g}$ (yellow), as well as s and p (cyan) orbitals of iron, respectively.}
% \label{fig:dosferrite}
% \end{figure}

% Suggestion for subfigure configuration. Might require additional LaTeX packages (subcaption). Update references for the individual subfigures.

The density of states for the two spin orientations are drawn in distinct directions. 
As expected, we observe a shift of the d-density of states due to the spin-dependent exchange-correlation potential. 
The Fermi level lies in a quasi-gap in the minority spin direction.

We introduced a stacked representation of the projected density of states.
In this representation, the area of a colored region represents a projection, rather than its function value.
It provides an intuitive impression of important contributions, but is less suitable for quantitative comparisons.

\paragraph{Unit cell optimization:}
In solids, the atomic structure is given by the positions of all atoms in the unit cell and the three primitive lattice vectors. 
The unit cell lattice vectors can be optimized using the stress tensor within a so-called cell dynamics.\cite{parrinello80_prl45_1196, parrinello1981polymorphic} 

In order to optimize only the lattice constant, we include the following constraint in the control input file:
\begin{verbatim}
!CELL
  FRIC=0.01 CONTRAINTTYPE='ISOTROPIC'
!END
\end{verbatim}
The parameter \verb|CONTRAINTTYPE='ISOTROPIC'| suppresses shear and uniaxial distortions. 
The cell dynamics shall be enabled only after the wave functions and the internal atomic positions have been optimized.

Once the calculation is completed, the last volume in the calculation is retrieved with 
%{\scriptsize
\begin{verbatim}
paw_get -w volume -u AA^3 ferrite
\end{verbatim}
%}
that provides the equilibrium in units of \AA$^3$. Here, $a_{\text{lat}}=\sqrt[3]{2V}=2.834$~\AA, which is 
within 1\% of the experimental lattice constant of $a_{lat}=2.860$~{\AA}.\cite{seki05_isij45_1789}
In order to monitor the convergence, we need to inspect the protocol file via 
\begin{verbatim}
grep -i T1 ferrite.prot
\end{verbatim}
that collects all lines containing the first lattice vector.
%The lattice is only written out with every larger write-up.

Note that as the unit cell changes, the k-point grid scales with the unit cell vectors. 
As a result, the k-point density may deviate from the intended value. 
The k-point distribution can be adjusted by restarting the calculation with the final lattice vectors. 

%===========================================================
\paragraph{Hexaferrum:}
%===========================================================
For hexaferrum we proceed analogously to ferrite:

Hexaferrum is hexagonal with experimental lattice constants of $a=2.517$~{\AA}, $c=4.060$~{\AA} and $c/a=1.613$, respectively.\cite{marinelli01_zmetallkd92_489} 
For the sake of simplicity, we study the material without spin polarization and with $c/a=1.633$ of the ideal hcp lattice. 
The effects of spin distribution and of the c/a ratio have been extensively investigated by Steinle-Neumann et al.\cite{steinleneumann99_prb60_791} and Choi et al.\cite{choi22_materials15_1276}, respectively.

For the non-magnetic phase, we can work with \verb|!STRUCTURE!GENERIC:NSPIN=1|, as described above. 
The total spin must be zero, which can be set with 
\verb|!STRUCTURE!GENERIC:SPIN[HBAR]=0.|, which happens to be the default.

The resulting density of states is shown in Fig.~\ref{fig:dosferrite}.

%===========================================================
\paragraph{Scanning the energy versus volume:}
%===========================================================

In order to evaluate the transition pressure of a pressure-induced phase transition, it is necessary to map the energy $E_X(V)$ as a function of volume for the two phases $X\in\{\text{bcc, hcp}\}$.  
The slope of the common tangent provides the transition pressure. 

We begin here with ferrite in the bcc phase.
The energy versus volume curve can be built with the bash script \verb|paw_scanlat.sh|:

{\scriptsize
\begin{verbatim}
export ROOTNAME=ferrite
export NJOBS=10
export LIST='90 92 94 96 98 100 102 104 106 108 110'
paw_scanlat.sh -p ${ROOTNAME} -l ${LIST} -j ${NJOBS}
paw_scanlat.sh -u -p ${ROOTNAME} -l ${LIST}
\end{verbatim}
}

The variable "\verb|LIST|" holds the percentage values to adjust the length unit \verb|LUNIT[AA]| in the corresponding structure file. The variable \verb|NJOBS| sets the number of calculations allowed to run simultaneously, e.g. the number of available cores on the CPU.

%\verb|paw_scanlat| constructs a directory for each scalefactor in the variable \verb|LIST| and copies the control file, structure file, and restart file into each.
%The length unit \verb|LUNIT[AA]| is adjusted according to the percentage value in \verb|LIST|. Then it executes each project.  

The chosen grid of lattice constants is fairly coarse. 
This is done to control the well-known sawtooth behavior,\cite{yin85_proc27confphyssemicond} which is common to all plane-wave-based codes:
For a given plane-wave cutoff, the basis set increases with increasing lattice constant. 
Because the number of (augmented) plane waves changes abruptly, the total energy drops by a small amount.
These energy steps act like noise, which prohibits  interpolation by a smooth function. 
Yet, using a coarse grid, this noise is averaged out by the interpolation.

\begin{figure}[!hbt]
  \centering
  \begin{subfigure}{\linewidth}
    \centering
    \includegraphics[width=\linewidth]{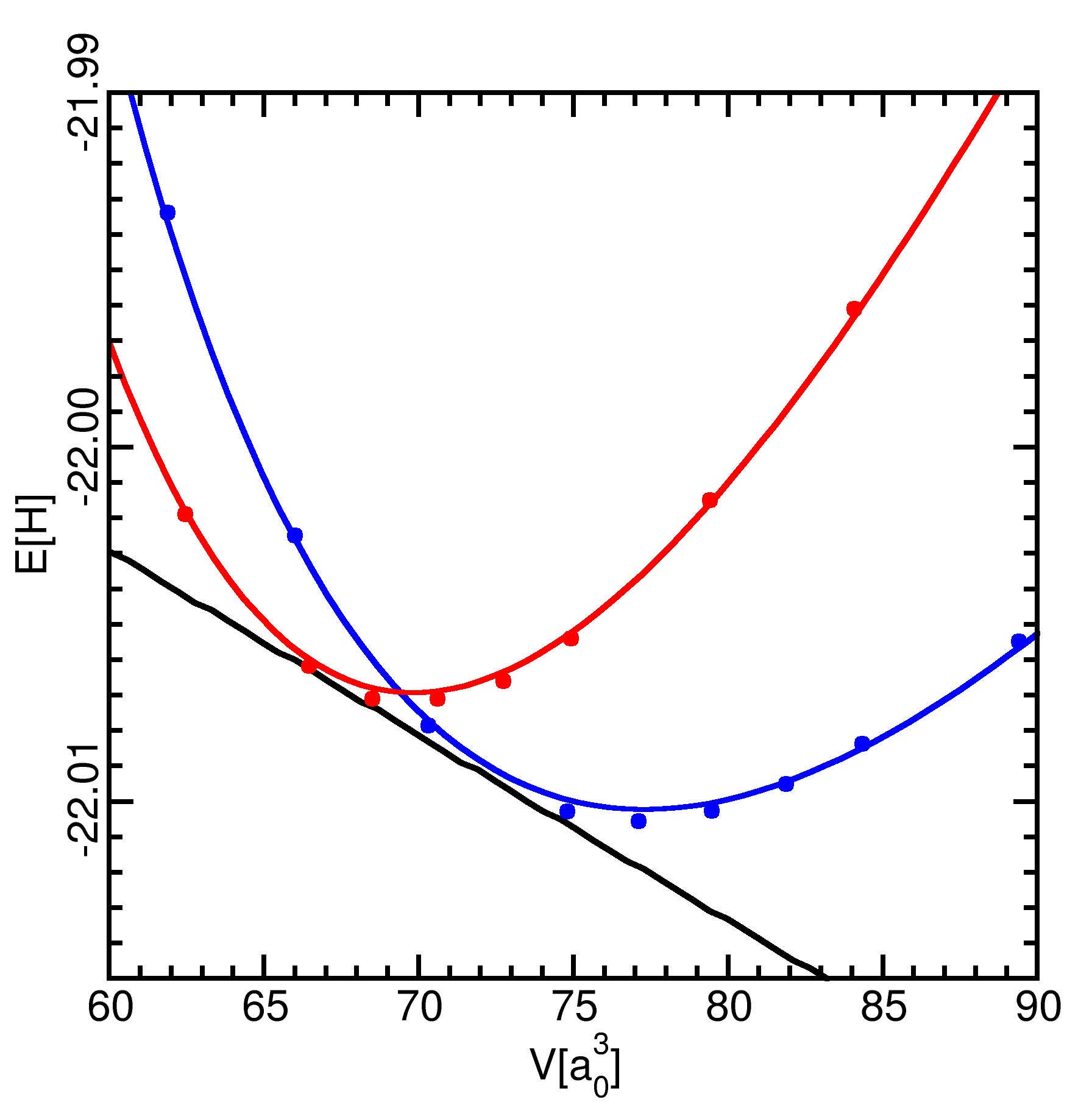}
    \caption{Energy versus volume. 
    The common tangent is identifying the pressure-induced transition.}
    \label{fig:eofv:a}
  \end{subfigure}
  \vspace{0.5em}
  \begin{subfigure}{\linewidth}
    \centering
    \includegraphics[width=\linewidth]{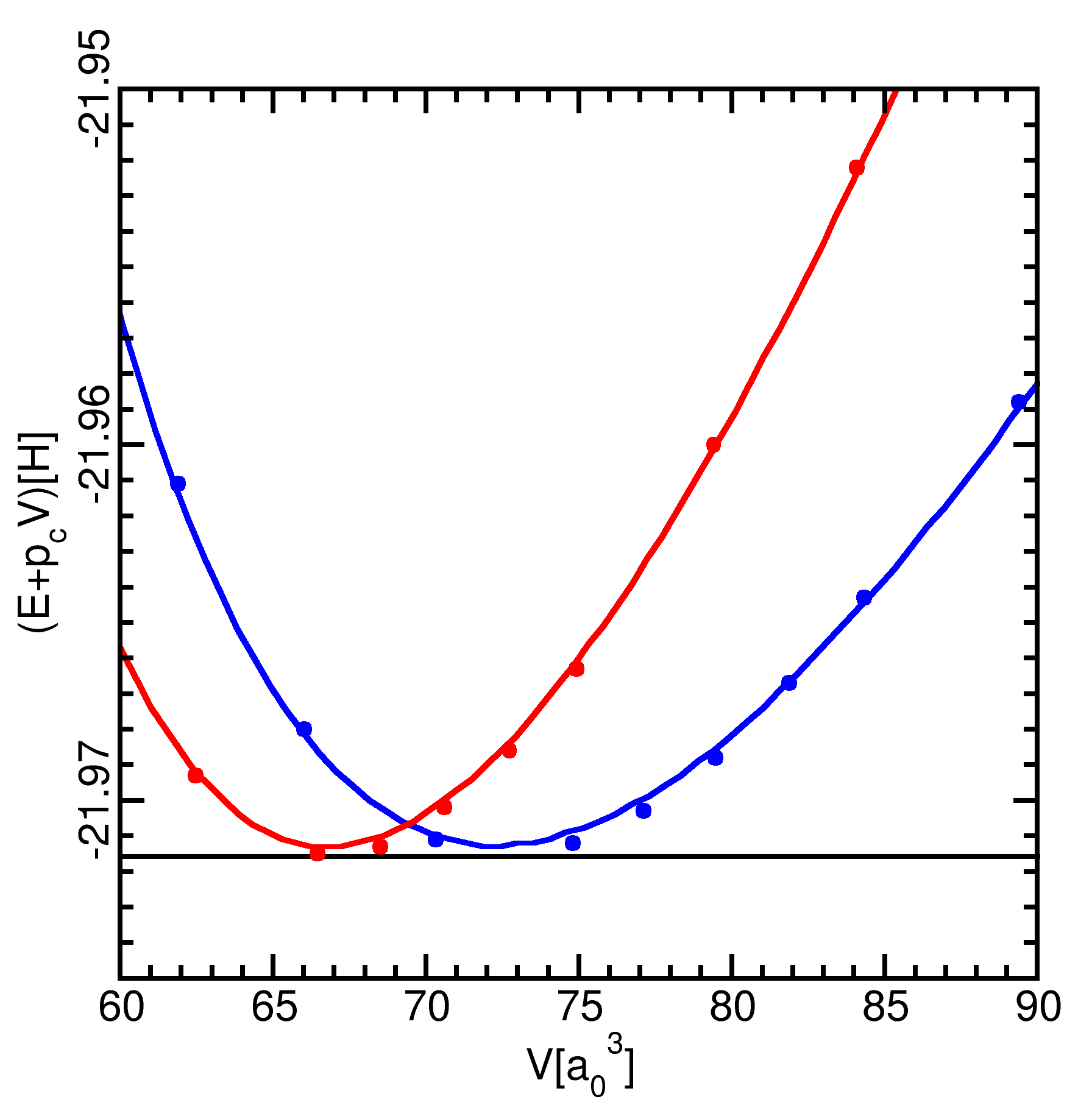}
    \caption{Energy including the $pV$ contribution of the reservoir at the transition pressure
    $p^* = 15.364$~GBar $= 0.522\times10^{-3}~H/a_B^3$.}
    \label{fig:eofv:b}
  \end{subfigure}
  \caption{\label{fig:eofv}
  Volume dependent energy curves for ferrite (blue) and hexaferrum (red).
  }
\end{figure}

With the option \verb|-u| of \verb|paw_scanlat|, a file latscan is produced, which holds a list of volumes and the corresponding energies. This file is passed into the \verb|paw_murnaghan| tool by
\begin{verbatim}
 paw_murnaghan -scale 1 <latscan   
\end{verbatim}
that interpolates the data to extract the equilibrium volume, bulk modulus, etc. via the well-known Murnaghan equation of state\cite{murnaghan44_pnas30_244} 
\begin{eqnarray}
    E(V)&=&E_0+\frac{B_0V_0}{B'}
    \Big[\frac{1}{B'-1}\left(\frac{V}{V_0}\right)^{1-B'}
\nonumber\\
    &&+\frac{V}{V_0}-1\Bigr],
    \label{eq:murnaghan}
\end{eqnarray}
which interpolates the energy $E$ as a function of volume $V$.
The fit parameters are the equilibrium energy $E_0$ and the associated equilibrium volume $V_0$, as well as the bulk modulus $B_0$ at the equilibrium volume and its pressure derivative $B'$. 
The \verb|paw_murnaghan| tool produces a file \verb|murn.dat|, which lists the interpolated $E(V)$ on a grid for inspection. The corresponding results, including the common tangent construction, whose slope yields the transition pressure, are plotted in Fig.~\ref{fig:eofv:a}. 

Figure~\ref{fig:eofv:b} shows the energy including the energy $pV$ of the volume reservoir at the transition pressure. The common minimum value is the enthalpy at the transition pressure.

%Also, the enthalpy $E(V)+pV$, where $p$ is the pressure and $V$ the associated cell volume, is shown, demonstrating that at the transition pressure, the enthalpy of the two phases is identical. 

% \begin{figure}[!hbt]
% \includegraphics[width=\linewidth]
% {Inputfiles/Peter2/Figs/Fig8/fig8a.png}
% %{Inputfiles/Peter/Figs/iron_eofv.png}
% %{data/fig8a.png}
% \includegraphics[width=\linewidth]{Inputfiles/Peter2/Figs/Fig8/fig8b.png}
% %{Inputfiles/Peter/Figs/iron_eofv-pv.png}
% %{data/fig8a.png}
% \caption{\label{fig:eofv}
% Energy in Hartree versus volume in $a_B^3$ for ferrite (blue) and hexaferrum (red).
% The common tangent, identifying the pressure induced transition is shown in black.
% The lower figure shows the energy including the energy $pV$ of the reservoir for the transition pressure $p^*=15.364$~GBar~$=0.522\times 10^{-3}~H/a_B^3$.
% }
% \end{figure}

% Suggestion for subfigure configuration. Might require additional LaTeX packages (subcaption). Update references for the individual subfigures.

The energy difference between hexaferrum and ferrite at the equilibrium volumes at zero pressure is 90~meV per atom, whereas the equilibrium volume of hexaferrum is 10\% smaller than that of ferrite, as obtained by Eq.~\ref{eq:murnaghan} and displayed in Table~\ref{tab:iron_volume} together with the corresponding bulk modulus. 

\begin{table}[hbt]
\begin{center}
\begin{tabular}{lcc}
\hline
\hline
    & ferrite & hexaferrum \\
\hline
$V_0$ & 11.45~\AA$^3$ & 10.35~\AA$^3$ \\
 & (11.38~\AA$^3$)\cite{nandi10_jpcm22_345501}
%&(10.26~\AA$^3$)\cite{choi22_materials15_1276}\\
% & (11.70~\AA$^3$)\cite{seki05_isij45_1789}& 
  % (11.14~\AA$^3$)\cite{marinelli01_zmetallkd92_489}\\&
 &(10.22~\AA$^3$)\cite{steinleneumann99_prb60_791}\\
 $B_0$ & 191 GPa & 307 GPa \\ 
&(204 GPa)\cite{nandi10_jpcm22_345501} &
% (192 GPa)\cite{choi22_materials15_1276}\\ & & 
 (292~GPa)\cite{steinleneumann99_prb60_791}\\
\hline
\hline
\end{tabular}
\end{center}
\caption{Lattice properties of ferrite and hexaferrum. For comparison, the corresponding results of DFT calculations of others are shown in parenthesis.
The data for hexaferrum are for the non-magnetic case and the ideal $c/a$ ratio of the hcp lattice.  
}
\label{tab:iron_volume}
\end{table}

\paragraph{Pressure-induced phase transition:}

The conventional way to determine pressure-induced phase transitions is to compute $E_j(V)$ on a grid for both phases and then search for the common tangent. 
Therein, $j\in\{A,B\}$, where $A$ and $B$ stands for the two relevant phases, ferrite and hexaferrum.

Nevertheless, we will proceed differently here by employing calculations at constant pressure, which have the advantage that not only the volume but also the shape of the unit cell can be optimized.

As before, the enthalpies of the two phases $j
\in\{A,B\}$ are
\begin{eqnarray}
    H_{j}(p)=\min_V\Bigl(E_j(V)+pV\Bigr).
\end{eqnarray}
The two phases coexist at the transition pressure $p^*$, for which the enthalpies of the two phases are equal, i.e. $H_{A}(p^*)=H_B(p^*)$. 
From these enthalpies, obtained for a given pressure $p$, an estimate of the transition pressure is computed by
\begin{eqnarray}
p^*\approx p-\frac{H_A(p)-H_B(p)}{V_A(p)-V_B(p)}.
\end{eqnarray}
The transition pressure is then obtained by iterating this equation to convergence.

The iteration is initiated with the energies at zero pressure obtained already, which in that case are identical to the enthalpies since $p=0$.
The pressure is specified in the \verb|!CELL| block of the control input file, i.e.
{\scriptsize
\begin{verbatim}
!CELL
  FRIC=0.01 CONSTRAINTTYPE='ISOTROPIC' P[GPa]=0.
!END
\end{verbatim}
}

We can specify the pressure in Hartree atomic units, or in the common unit GPa by using \verb|P[GPa]| rather than \verb|P| as keyword. 
The current values for the energy and volume can be obtained using \verb|paw_get -w etot -u H|~\textit{rootname}, as well as 
\verb|paw_get -w volume -u cubabohr|~\textit{rootname}, and are shown in Table~\ref{ta:transition_pressure}.
For a calculation at finite pressure, the energy returned is always the enthalpy including the $+pV$ term.
\begin{table}[!hbt]
\begin{center}
\begin{tabular}{lrrr}
\hline
\hline
& $a_{\text{lat}}$[\AA] & V[\AA$^2]$ & $H$\\
\hline
bcc P=0 & 2.834~\AA & 11.382~\AA$^3$ & -22.0106 $H$\\
bcc P=$p^*$& 2.780~\AA & 10.746~\AA$^3$ &-21.9716~$H$\\
hcp P=0 & 2.445~\AA &10.333~\AA$^3$ & -22,0072~$H$ \\
hcp P=$p^*$ & 2.410~\AA &9.903 \AA$^3$ & -21.9716~$H$\\
\hline
\hline
    \end{tabular}
\end{center}
\caption{Lattice constants, equilibrium volumes and enthalpies for zero pressure and for the calculated transition pressure 
$p^*=15.364$~GPa of the ferrite to hexaferrum phase transition.
For hexaferrum we specify $a_{\text{lat}}=a$ and use the ideal $c/a$ ratio of the hcp phase.
}
\label{ta:transition_pressure}
\end{table}

%Ferrite
%P=0; alat=2*1.417074*AA
%P=15.36381523793948595587 GPa alat= 2* 1.390171 \AA

%Hexaferrum
%P=0   alat=2.444859 AA clat=3.992422AA
%P=15.364 GBar=5.22\times 10^{-4}a.u..  alat=2.410399 clat=3.936166

%=====================================================
\subsection{Exercise 4: Correlated oxide}
%=====================================================
%\petertt{Warning: Praseodymium setup is close to the overcompleteness problem NPRO=2 1 1 1 works but NPRO=2 2 1 1 fails for structure optimization.}

Praseodymium-manganite, or PrMnO$_3$, is one member of a class of magnetic materials with strong correlations between electron, spin, and phonon degrees of freedom. 
Manganites typically form polaron solids and polaron liquids with an extremely rich phase diagram.\cite{martin1999magnetic} 

In this exercise, we will become familiar with antiferromagnets, which require special treatment.
We will use the local hybrid functional PBE0r, described in Section~\ref{sec:local_hybrid_functional}, which can be regarded as a range-separated hybrid functional,
where the cutoff for the exchange interactions is defined by localized tight-binding orbitals. This allows us to study materials with strong correlations for which conventional density functionals fail.
Furthermore, we will become familiar with f-electron systems such as Pr.

\paragraph{Antiferromagnetic order:}
First, we take antiferromagnetic ordering into account. 
Materials with high-spin atoms or d- or f-electrons in the valence require special treatment because their energy surface typically exhibits many metastable minima corresponding to distinct antiferromagnetic orders.

The barriers between these metastable states are very high, often several eV. 
An atom can change its spin orientation only by passing through an unfavorable low-spin configuration.
This problem may be avoided by allowing non-collinear spin distributions, where the spins may rotate.
However, the typical use case is to target specific antiferromagnetic orders.

For that purpose, we bias the system so that it falls into a basin with the desired magnetic order.
The bias is removed as soon as the system has reached the desired basin.
The bias is applied in the structure file by including an \verb|!STRUCTURE!ORBPOT| block, as shown below:
{\footnotesize
\begin{verbatim}
!ORBPOT
  !POT ATOM='MN5' VALUE=+.2  TYPE='D'  S=1 !END
  !POT ATOM='MN6' VALUE=+.2  TYPE='D'  S=1 !END
  !POT ATOM='MN7' VALUE=-.2  TYPE='D'  S=1 !END
  !POT ATOM='MN8' VALUE=-.2  TYPE='D'  S=1 !END

  !POT ATOM='PR1' VALUE=-.15 TYPE='F1' S=1 !END
  !POT ATOM='PR1' VALUE=-.15 TYPE='F7' S=1 !END
  !POT ATOM='PR2' VALUE=+.15 TYPE='F1' S=1 !END
  !POT ATOM='PR2' VALUE=+.15 TYPE='F7' S=1 !END
  !POT ATOM='PR3' VALUE=-.15 TYPE='F1' S=1 !END
  !POT ATOM='PR3' VALUE=-.15 TYPE='F7' S=1 !END
  !POT ATOM='PR4' VALUE=+.15 TYPE='F1' S=1 !END
  !POT ATOM='PR4' VALUE=+.15 TYPE='F7' S=1 !END
!END
\end{verbatim}
}
This adds a site-dependent, angular-momentum-dependent, and spin-dependent potential to the wave functions. 
With \verb|ATOM=|, a particular atom is specified, whereas by \verb|TYPE='D'| we select all orbitals with $\ell=2$. 
With $S=1$, we require that, in a collinear calculation, the potential acts along the spin axis. 
The sign of \verb|VALUE=| determines the orientation of the spin potential.
The spins of the Mn ions are oriented to form antiferromagnetically coupled (001) planes, which is the experimentally observed order.
The first four lines give a bias for the MN5 and MN6 atoms in the spin-down direction and for the atoms MN7 and MN8 in the spin-up direction.

In addition to the spin orientation of the Mn-ions, we also specify the orientation of the f-electrons on the Pr-ions. 
Here, we not only specify the main angular momentum, but we also choose certain orbitals 'F1' and 'F7'. The spin-orientation of the f-electrons is less relevant; however, as a consequence of using hybrid functionals, we need to guide the f-electron atoms into a high-spin configuration and the correct oxidation state.

However, it is important to disable the \verb|!ORBPOT| branch before completing the convergence, as detailed below.

\paragraph{Local PBE0r hybrid functional:}

%\thomastt{Ich werde hier noch Details zum PBE0r XC Funktional einfügen}

In order to activate our local hybrid functional, we include a branch
\begin{verbatim}
!DFT !NTBO !END !END 
\end{verbatim}
in the control input file.
In addition, we need to set some data for each atom species in each
\verb|!STRUCTURE!SPECIES|. For the Mn ions, for instance, we choose
\begin{verbatim}
!NTBO NOFL=1 0 1 LHFWEIGHT=0.07 !END
\end{verbatim}

The first parameter \verb|NOFL=| specifies the number of local orbitals for each angular momentum, i.e. here for $\ell=0,1,2,\ldots$. 
As a guideline, one should introduce $1$ for each partially occupied angular-momentum shell. The value $2$ shall be used in rare occasions, when two shells with the same angular momentum are treated outside the frozen-core approximation, and both are occupied simultaneously. 
Otherwise, the value is set to zero.

The most important parameter is \verb|LHFWEIGHT|, which is a scale factor defined as $a_R$ in Eq.~\ref{eq:PBE0r}. 
This is an empirical parameter for each atom species. 
A value in the range $0.07-0.1$, corresponding to 7-10\% of exact exchange, is a good choice for most orbitals.

\paragraph{Wave function optimization:}
We first optimize the wave functions with the spin-bias for a few, e.g. 100 iterations.

%\petertt{Thomas, please read the following:}
During these initial iterations, we set the plane-wave cutoff to $E_{PW}=15$~Ry, which is a relatively low value.
This is a cure for a common problem during the initial steps of the wave function optimization, namely that the dynamics of the electronic degrees of freedom becomes too fast for the orthogonality constraint to be enforced.

After the iterations with spin bias, it may be a good idea to run the \verb|paw_dos| tool once to verify that the correct magnetic order has been established. 
The protocol file of the \verb|paw_dos| tool provides a list of the spin moments of all atoms.
Once the antiferromagnetic order has been verified, we disable the block \verb|!STRUCTURE!ORBPOT| in the structure input file and continue the wave function optimization.
Do not forget to set \verb|!CONTROL!GENERIC:START=F|.
After the wave functions are converged, we optimize the atomic structure.

\begin{figure}[!hbt]
    \centering
    \includegraphics[width=\linewidth]
    {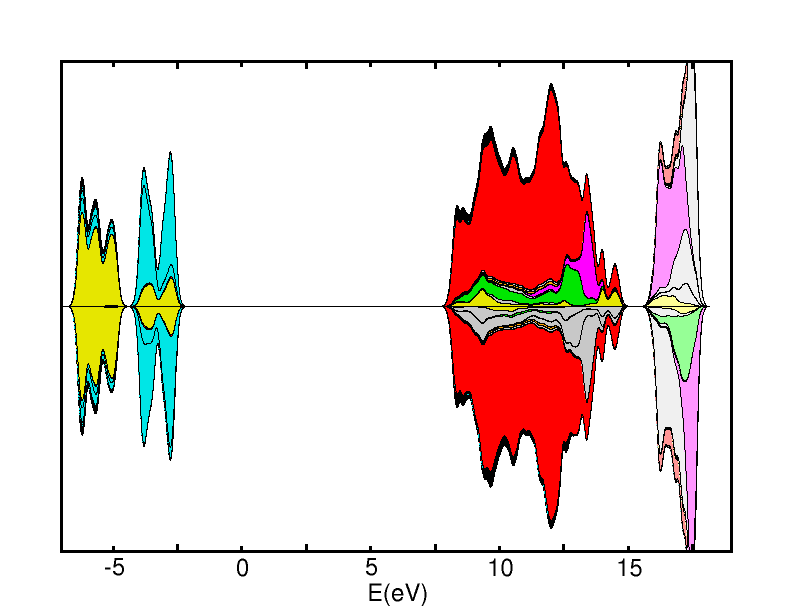}
    %{Inputfiles/Peter/Figs/pmo_dos.png}
    %{data/fig9.png}
    \caption{Projected density of states of PrMnO$_3$. 
    The projected density of states are stacked on top of each other so that the color rather than the height determines the weight. 
    The density of spin-up and spin-down states are drawn in opposite directions.
    Below 0~eV, we find the oxygen s-orbitals (yellow) and the Pr-p states (cyan).
    The valence band is dominated by the O-p states (red). 
    The Mn-$e_g$ states are in yellow and the Mn-$t_{2g}$ states are shown in light green, whereas the Pr-f states are shown in magenta. 
    The density of states of atoms with a positive majority-spin direction is colored, while those with a downward majority-spin direction are shaded in grey.}
    \label{fig:pmo_dos}
\end{figure}

\paragraph{Analysis:}
For the density of states we distinguish atoms with different spin orientation.
%\thomastt{The color code seems to be inconsistent with Fig. 9}
The main feature, shown in Fig.~\ref{fig:pmo_dos}, are the oxygen valence band with O-p character (red). Below the O-p band, we see the oxygen s-band (yellow) and the Pr-p states (cyan). The Pr-f states (magenta) are split into filled and empty multiplets. 
This splitting is due to the large Coulomb repulsion between the f-electrons, which is included through the explicit
exchange term in the hybrid functional.

Relevant for the polaron order are the Mn-d orbitals. 
The three $t_{2g}$ orbitals of each Mn atom are fully spin aligned, reflecting Hund's rule. 
The two $e_g$ orbitals of each Mn atom, having the same spin as the $t_{2g}$ orbitals, are furthermore split into a filled lower band and an empty upper band. 
This splitting reflects the orbital polarization resulting from the Jahn-Teller effect.

The polaron formation can be seen from the output of the structure tool \verb|paw_strc|. 
It prints the bond lengths and bond angles for each atom, respectively.
Inspecting the nearest neighbors of an Mn-ion, one 
recognizes the Jahn-Teller effect responsible for the polaron order in PrMnO$_3$. 
Each Mn-site has short bonds and two long bonds in the ferromagnetic (001)-plane, and two intermediate bonds perpendicular to the (001)-planes.  
The elongation of the long bond is caused by the $e_g$ electron in the antibond with the same orientation. 
The elongation lifts the degeneracy of the two $e_g$ electrons in the majority spin direction.

\section{Data Availability}

The data underlying this study are openly available at \verb|https://github.com/cp-paw/tutorial|

\section{Acknowledgement}
%\begin{acknowledgement}
This work is funded in parts by the Deutsche Forschungsgemeinschaft (DFG, German Research Foundation) via project number 217133147/SFB1073 (projects B03 and C03), 417590517/SFB1415 and 519869949.
%\end{acknowledgement}

%\begin{suppinfo}

%This will usually read something like: ``Experimental procedures and characterization data for all new compounds. The class will automatically add a sentence pointing to the information on-line:

%\end{suppinfo}

\appendix

%=======================================================
\section{Augmentation parameters}
\label{sec:setupfiles}
%=======================================================
As mentioned above, the CP-PAW code defines the augmentation on-the-fly during the initialization phase of each calculation.
It requires a set of parameters, which may not be trivial to choose by the novice. 
Therefore, we supply parameters for the relevant elements:

\begin{itemize}
\item \verb|h_.stp| {\scriptsize\verbatiminput{Inputfiles/Peter2/src/Setups/h_.stp}}
\item \verb|c_.stp| {\scriptsize\verbatiminput{Inputfiles/Peter2/src/Setups/c_.stp}}
\item \verb|o.stp| {\scriptsize\verbatiminput{Inputfiles/Peter2/src/Setups/o_.stp}}
\item \verb|mn.stp| {\scriptsize\verbatiminput{Inputfiles/Peter2/src/Setups/mn.stp}}
\item \verb|fe.stp| {\scriptsize\verbatiminput{Inputfiles/Peter2/src/Setups/fe.stp}}
\item \verb|pr.stp| {\scriptsize\verbatiminput{Inputfiles/Peter2/src/Setups/pr.stp}}
\end{itemize}

\bibliography{all}

%\bibliographystyle{plain}
%\bibliography{all}
\end{document}